# High-dimensional Linear Discriminant Analysis: Optimality, Adaptive Algorithm, and Missing Data[1]

T. Tony Cai and Linjun Zhang

University of Pennsylvania

**Abstract**

This paper aims to develop an optimality theory for linear discriminant analysis in the high-dimensional setting. A data-driven and tuning free classification rule, which is based on an adaptive constrained $\ell_1$ minimization approach, is proposed and analyzed. Minimax lower bounds are obtained and this classification rule is shown to be simultaneously rate optimal over a collection of parameter spaces.

In addition, we consider classification with incomplete data under the missing completely at random (MCR) model. An adaptive classifier with theoretical guarantees is introduced and optimal rate of convergence for high-dimensional linear discriminant analysis under the MCR model is established. The technical analysis for the case of missing data is much more challenging than that for the complete data. We establish a large deviation result for the generalized sample covariance matrix, which serves as a key technical tool and can be of independent interest. An application to lung cancer and leukemia studies is also discussed.

## 1 Introduction

Classification is one of the most important tasks in statistics and machine learning with applications in a broad range of fields. See, for example, Hastie et al. (2009). The problem has been well studied in the low-dimensional setting. In particular, consider the Gaussian case where one wishes to classify a new random vector $\boldsymbol{Z}$ drawn with equal probability from one of two Gaussian distributions $N_p(\boldsymbol{\mu}_1, \Sigma)$ (class 1) and $N_p(\boldsymbol{\mu}_2, \Sigma)$ (class 2). In the ideal setting where all the parameters $\boldsymbol{\theta} = (\boldsymbol{\mu}_1, \boldsymbol{\mu}_2, \Sigma)$ are known, Fisher's linear discriminant rule, which is given by

$$C_{\boldsymbol{\theta}}(\boldsymbol{Z}) = \begin{cases} 1, & \boldsymbol{\delta}^\top \Omega (\boldsymbol{Z} - \frac{\boldsymbol{\mu}_1 + \boldsymbol{\mu}_2}{2}) < 0 \\ 2, & \boldsymbol{\delta}^\top \Omega (\boldsymbol{Z} - \frac{\boldsymbol{\mu}_1 + \boldsymbol{\mu}_2}{2}) \geq 0, \end{cases} \qquad (1)$$

where $\boldsymbol{\delta} = \boldsymbol{\mu}_2 - \boldsymbol{\mu}_1$, and $\Omega = \Sigma^{-1}$ is the precision matrix, is well known to be optimal (Anderson, 2003). Fisher's rule separates the two classes by a linear combination of features and its misclassification error is $R_{\mathrm{opt}}(\boldsymbol{\theta}) = \Phi\left(-\frac{1}{2}\Delta\right)$, where $\Phi$ is the cumulative distribution function of the standard normal distribution and $\Delta = \sqrt{\boldsymbol{\delta}^\top \Omega \boldsymbol{\delta}}$ is the signal-to-noise ratio.

[1]Department of Statistics, The Wharton School, University of Pennsylvania, Philadelphia, PA 19104. The research was supported in part by NSF Grant DMS-1712735 and NIH Grant R01 GM-123056.

Although Fisher's rule can serve as a useful performance benchmark, it is not practical for real data analysis as the parameters $\boldsymbol{\mu}_1, \boldsymbol{\mu}_2$ and $\Sigma$ are typically unknown and need to be estimated from the data. In applications, it is desirable to construct a data-driven classification rule based on two observed random samples, $\boldsymbol{X}_1^{(1)}, ..., \boldsymbol{X}_{n_1}^{(1)} \overset{i.i.d.}{\sim} N_p(\boldsymbol{\mu}_1, \Sigma)$ and $\boldsymbol{X}_1^{(2)}, ..., \boldsymbol{X}_{n_2}^{(2)} \overset{i.i.d.}{\sim} N_p(\boldsymbol{\mu}_2, \Sigma)$. In the conventional low-dimensional setting, this is easily achieved by plugging in Fisher's linear discriminant rule (1) the corresponding sample means and pooled sample covariance matrix for the parameters $\boldsymbol{\mu}_1, \boldsymbol{\mu}_2$ and $\Sigma$ respectively. This classification rule is asymptotically optimal when the dimension $p$ is fixed. See, for example, Anderson (2003).

Driven by many contemporary applications, much recent attention has been on the high-dimensional setting where the dimension is much larger than the sample size. In this case, the sample covariance matrix is not even invertible and it is difficult to estimate the precision matrix $\Omega$. The standard linear discriminant rule thus fails completely. Several regularized classification methods, including the regularized logistic regression (Shevade and Keerthi, 2003), Naive Bayes method (Bickel and Levina, 2004), hard thresholding (Shao et al., 2011), direct estimation methods in (Cai and Liu, 2011; Mai et al., 2012), have been proposed for classification of high-dimensional data. In particular, Cai and Liu (2011) introduced a direct estimation method for the high-dimensional linear discriminant analysis based on the key observation that the ideal Fisher's discriminant rule given in (1) depends on the parameters $\boldsymbol{\mu}_1, \boldsymbol{\mu}_2$ and $\Sigma$ only through the discriminant direction $\boldsymbol{\beta} = \Omega \boldsymbol{\delta}$. They proposed to estimate the discriminant direction $\boldsymbol{\beta}$ directly instead of estimating $\Sigma$ and $\boldsymbol{\delta}$ separately, under the assumption that $\boldsymbol{\beta}$ is sparse. It was shown that their classification rule is consistent.

Despite much recent progress in methodological development on high-dimensional classification problems, there has been relatively little fundamental study on the optimality theory for the discriminant analysis. Minimax study of high-dimensional discriminant analysis has been considered in Azizyan et al. (2013) and Li et al. (2017) in the special case where the covariance matrix $\Sigma = \sigma^2 I$ for some $\sigma > 0$. However, even in this relatively simple setting there is still a gap between the minimax upper and lower bounds. It is unclear what the optimal rate of convergence for the minimax misclassification risk is and which classification rule is rate optimal under the general Gaussian distribution. The first major goal of the present paper is to provide answers to these questions. Furthermore, although the problem of missing data arises frequently in the analysis of high-dimensional data, compared to the conventional low-dimensional setting, there is a paucity of methods for inference with incomplete high-dimensional data. The second goal of this paper is to develop an optimality theory for high-dimensional discriminant analysis with incomplete data and to construct in this setting a data-driven adaptive classifier with theoretical guarantees.

Given two random samples, $\boldsymbol{X}_1^{(1)}, ..., \boldsymbol{X}_{n_1}^{(1)} \overset{i.i.d.}{\sim} N_p(\boldsymbol{\mu}_1, \Sigma)$ and $\boldsymbol{X}_1^{(2)}, ..., \boldsymbol{X}_{n_2}^{(2)} \overset{i.i.d.}{\sim} N_p(\boldsymbol{\mu}_2, \Sigma)$, we wish to construct a classifier $\hat{C}$ to classify a future data point $\boldsymbol{Z}$ drawn from these two distributions with equal prior probabilities, into one of the two classes. Given the observed data, the performance of the classification rule is measured by the misclassification error

$$R_{\boldsymbol{\theta}}(\hat{C}) = \mathbb{P}_{\boldsymbol{\theta}}(\text{label}(\boldsymbol{Z}) \neq \hat{C}(\boldsymbol{Z})), \qquad (2)$$

where $\boldsymbol{\theta} = (\boldsymbol{\mu}_1, \boldsymbol{\mu}_2, \Sigma)$, $\mathbb{P}_{\boldsymbol{\theta}}$ denotes the probability with respect to $\boldsymbol{Z} \sim \frac{1}{2} N_p(\boldsymbol{\mu}_1, \Sigma) + \frac{1}{2} N_p(\boldsymbol{\mu}_2, \Sigma)$ and $\boldsymbol{Z}$ is independent of the observed $\boldsymbol{X}$'s. label($\boldsymbol{Z}$) denotes the true class of $\boldsymbol{Z}$. For a given classifier $\hat{C}$, we use the excess misclassification risk relative to the oracle rule (1), $R_{\boldsymbol{\theta}}(\hat{C}) - R_{\text{opt}}(\boldsymbol{\theta})$, to measure the performance of the classifier $\hat{C}$. Let $n = \min\{n_1, n_2\}$. We consider in this paper a collection of the parameter spaces $\mathcal{G}(s, M_{n,p})$ defined by

$$\mathcal{G}(s, M_{n,p}) = \{\boldsymbol{\theta} = (\boldsymbol{\mu}_1, \boldsymbol{\mu}_2, \Sigma) : \boldsymbol{\mu}_1, \boldsymbol{\mu}_2 \in \mathbb{R}^p, \Sigma \in \mathbb{R}^{p \times p}, \Sigma \succ 0,$$
$$\|\boldsymbol{\beta}\|_0 \leq s, M^{-1} \leq \lambda_{\min}(\Sigma) \leq \lambda_{\max}(\Sigma) \leq M, M_{n,p} \leq \Delta \leq 3 M_{n,p}\}, \qquad (3)$$

where $M > 1$ is a constant, $M_{n,p} > 0$ can potentially grow with $n$ and $p$, and $\lambda_{\max}(\Sigma)$ and $\lambda_{\min}(\Sigma)$ are respectively the largest and smallest eigenvalue of $\Sigma$. The notation $\Sigma \succ 0$ means that $\Sigma$ is symmetric and positive definite. Recall that $\Delta = \sqrt{\boldsymbol{\delta}^\top \Omega \boldsymbol{\delta}}$ and $\boldsymbol{\beta} = \Omega \boldsymbol{\delta}$.

Combining the upper and lower bounds results given in Section 3 leads to the following minimax rates of convergence for the excess misclassification risk.

**Theorem 1.** *Consider the parameter space $\mathcal{G}(s, M_{n,p})$, $s$ and $p$ approach infinity as $n$ grows to infinity, and $M_{n,p}\sqrt{\frac{s \log p}{n}} = o(1)$ with $n \to \infty$,*

1. *If $M_{n,p}$ is a fixed constant not depending on $n$ and $p$, then for any constant $\alpha > 0$, there exist two constants $C_\alpha^{(2)} > C_\alpha^{(1)} > 0$, such that*

$$\inf_{\hat{C}} \sup_{\boldsymbol{\theta} \in \mathcal{G}(s, M_{n,p})} \mathbb{P}\left(C_\alpha^{(1)} \cdot \frac{s \log p}{n} < R_{\boldsymbol{\theta}}(\hat{C}) - R_{\text{opt}}(\boldsymbol{\theta}) < C_\alpha^{(2)} \cdot \frac{s \log p}{n}\right) \geq 1 - \alpha.$$

2. If $M_{n,p} \to \infty$ as $n \to \infty$, then these exists a sequence $\delta_n$ with $\lim_{n\to\infty} \delta_n = 0$, such that for any constant $\alpha > 0$, there exist two constants $C_\alpha^{(2)} > C_\alpha^{(1)} > 0$ satisfying, for sufficiently large $n$,

$$\inf_{\hat{C}} \sup_{\boldsymbol{\theta} \in \mathcal{G}(s, M_{n,p})} \mathbb{P}\left(C_\alpha^{(1)} \frac{s \log p}{n} e^{-(\frac{1}{8}+\delta_n)M_{n,p}^2} < R_{\boldsymbol{\theta}}(\hat{C}) - R_{\text{opt}}(\boldsymbol{\theta}) < C_\alpha^{(2)} \frac{s \log p}{n} e^{-(\frac{1}{8}-\delta_n)M_{n,p}^2}\right) \geq 1-\alpha.$$

It is worth noting that $M_{n,p}$ controls the magnitude of $\Delta$, which is interpreted as the signal-to-noise ratio. As shown in the second case, when the signal-to-noise ratio grows, the classification problem becomes easier and our result precisely characterizes that the convergence rate is exponentially faster with an additional factor $\exp\left(-\left(1/8 + o(1)\right) M_{n,p}^2\right)$.

Furthermore, we propose a three-step data-driven classification rule, called *AdaLDA*, by using an adaptive constrained $\ell_1$ minimization approach which takes into account the variability of individual entries. This classification rule is shown to be simultaneously rate optimal over the collection of parameter spaces $\mathcal{G}(s, M_{n,p})$. To the best of our knowledge, this is the first optimality result for classification of high-dimensional Gaussian data. Furthermore, in contrast to many classification rules proposed in the literature, which require to choose tuning parameters, this procedure is data-driven and tuning free.

In addition, we also consider classification in the presence of missing data. As in the conventional low-dimensional setting, the problem of missing data also arises frequently in the analysis of high-dimensional data from in a range of fields such as genomics, epidemiology, engineering, and social sciences (Graham, 2009; Libbrecht and Noble, 2015; White et al., 2011). Compared to the low-dimensional setting, there are relatively few inferential methods for missing data in the high-dimensional setting. Examples include high-dimensional linear regression (Loh and Wainwright, 2012), sparse principal component analysis (Lounici, 2013), covariance matrix estimation (Cai and Zhang, 2016), and vector autoregressive (VAR) processes (Rao et al., 2017). In this paper, following the missing mechanism considered in the aforementioned papers, we investigate high-dimensional discriminant analysis in the presence of missing observations under the missing completely at random (MCR) model.

We construct a data-driven adaptive classifier with theoretical guarantees based on incomplete data and also develop an optimality theory for high-dimensional linear discriminant analysis under the MCR model. The technical analysis for the case of missing data is much more challenging than that for the complete data, although the classification procedure and the resulting convergence rates look similar. To facilitate the theoretical analysis, we establish a key technical tool, which is a large deviation result for the generalized sample covariance matrix. This is related to the masked covariance matrix estimator considered in Levina and Vershynin (2012) and Chen et al. (2012), see further discussions in Section 2.3.

This technical tool can be of independent interest as it is potentially useful for other related problems in high-dimensional statistical inference with missing data.

The proposed adaptive classification algorithms can be cast as linear programs and are thus easy to implement. Simulation studies are carried out to investigate the numerical performance of the classification rules. The results show that the proposed classifiers enjoy superior finite sample performance in comparison to existing methods for high-dimensional linear discriminant analysis. The proposed classifiers are also illustrated through an application to the analysis of lung cancer and leukemia datasets. The results show that they outperform existing methods.

The rest of the paper is organized as follows. In Section 2, after basic notation and definitions are reviewed, we introduce an adaptive algorithm for high-dimensional discriminant analysis with the complete data and then propose a more general procedure for the setting of incomplete data. Section 3 studies the theoretical properties of these classification rules and related estimators. In addition, minimax lower bounds are given. The upper and lower bounds together establish the optimal rates of convergence for the minimax misclassification risk. Numerical performance of the classification rules are investigated in Section 4 and the proofs of the main results are given in Section 5. Technical lemmas are proved in the Supplementary Material (Cai and Zhang, 2018).

## 2 Methodology

In this section, we firstly introduce an adaptive algorithm for high-dimensional linear discriminant analysis with the complete data. This algorithm is called AdaLDA (**Ada**ptive **L**inear **D**iscriminant **A**nalysis rule). We then propose a data-driven classifier, called ADAM (**A**daptive linear **D**iscriminant **A**nalysis with randomly **M**issing data), for the incomplete data under the MCR model.

### 2.1 Notation and definitions

We begin with basic notation and definitions. Throughout the paper, vectors are denoted by boldface letters. For a vector $\boldsymbol{x} \in \mathbb{R}^p$, the usual vector $\ell_0, \ell_1, \ell_2$ and $\ell_\infty$ norms are denoted respectively by $\|\boldsymbol{x}\|_0, \|\boldsymbol{x}\|_1, \|\boldsymbol{x}\|_2$ and $\|\boldsymbol{x}\|_\infty$. Here the $\ell_0$ norm counts the number of nonzero entries in a vector. The support of a vector $\boldsymbol{x}$ is denoted by $\text{supp}(\boldsymbol{x})$. The symbol $\circ$ denotes the Hadamard product. For $p \in \mathbb{N}$, $[p]$ denotes the set $\{1, 2, ..., p\}$. For $j \in [p]$, denote by $\boldsymbol{e}_j$ the $j$-th canonical basis in $\mathbb{R}^p$. For a matrix $\Sigma = (\sigma_{ij})_{1\leq i,j\leq p} \in \mathbb{R}^{p\times p}$, the Frobenius norm is defined as $\|\Sigma\|_F = \sqrt{\sum_{i,j} \sigma_{ij}^2}$ and the spectral norm is defined to be $\|\Sigma\|_2 = \sup_{\|\boldsymbol{x}\|_2=1} \|\Sigma\boldsymbol{x}\|_2$. The vector $\ell_\infty$ norm of the matrix $\Sigma$ is $|\Sigma|_\infty = \max_{i,j} |\sigma_{ij}|$. For

a symmetric matrix $\Sigma$, we use $\lambda_{\max}(\Sigma)$ and $\lambda_{\min}(\Sigma)$ to denote respectively the largest and smallest eigenvalue of $\Sigma$. $\Sigma \succ 0$ means that $\Sigma$ is positive definite. For a positive integer $s < p$, let $\Gamma(s) = \{\boldsymbol{u} \in \mathbb{R}^p : \|\boldsymbol{u}_{S^C}\|_1 \leq \|\boldsymbol{u}_S\|_1,$ for some $S \subset [p]$ with $|S| = s\}$, where $\boldsymbol{u}_S$ denotes the subvector of $\boldsymbol{u}$ confined to $S$. For two sequences of positive numbers $a_n$ and $b_n$, $a_n \lesssim b_n$ means that for some constant $c > 0$, $a_n \leq cb_n$ for all $n$, and $a_n \asymp b_n$ if $a_n \lesssim b_n$ and $b_n \lesssim a_n$. We say an event $\mathcal{A}_n$ holds with high probability if $\liminf_{n \to \infty} \mathbb{P}(\mathcal{A}_n) = 1$. Finally, $c_0, c_1, c_2, C, C_1, C_2, \ldots$ denote generic positive constants that may vary from place to place.

The complete data $\boldsymbol{X}_1^{(1)}, \ldots, \boldsymbol{X}_{n_1}^{(1)}$ and $\boldsymbol{X}_1^{(2)}, \ldots, \boldsymbol{X}_{n_2}^{(2)}$ are independent realizations of $\boldsymbol{X}^{(1)} \sim N_p(\boldsymbol{\mu}_1, \Sigma)$ and $\boldsymbol{X}^{(2)} \sim N_p(\boldsymbol{\mu}_2, \Sigma)$. We assume $n_1 \asymp n_2$ and define $n = \min\{n_1, n_2\}$. In our asymptotic framework, we let $n$ be the driving asymptotic parameter, $s$ and $p$ approach infinity as $n$ grows to infinity. The missing completely at random (MCR) model assumes that one observes samples $\{\boldsymbol{X}_1^{(1)}, \ldots, \boldsymbol{X}_{n_1}^{(1)}\}$ and $\{\boldsymbol{X}_1^{(2)}, \ldots, \boldsymbol{X}_{n_2}^{(2)}\}$ with missing values, where the observed coordinates of $\boldsymbol{X}_t^{(k)}$ are indicated by an independent vector $\boldsymbol{S}_t^{(k)} \in \{0,1\}^p$ for $t = 1, \ldots, n_k$, $k = 1, 2$, that is,

$$X_{tj}^{(k)} \text{ is observed if } S_{tj}^{(k)} = 1 \text{ and } X_{tj}^{(k)} \text{ is missing if } S_{tj}^{(k)} = 0; \ t \in [n_k], j \in [p], k = 1, 2. \quad (4)$$

Here $X_{tj}^{(k)}$ and $S_{tj}^{(k)}$ are respectively the $j$-th coordinate of the vectors $\boldsymbol{X}_t^{(k)}$ and $\boldsymbol{S}_t^{(k)}$. Generally, we use the superscript "$*$" to denote objects related to missing values. The incomplete samples with missing values are denoted by $\boldsymbol{X}^{(1)*} = \{\boldsymbol{X}_1^{(1)*}, \ldots, \boldsymbol{X}_{n_1}^{(1)*}\}$ and $\boldsymbol{X}^{(2)*} = \{\boldsymbol{X}_1^{(2)*}, \ldots, \boldsymbol{X}_{n_2}^{(2)*}\}$.

Regarding the mechanism for missingness, the MCR model is formally stated as below. This assumption is more general than the one considered previously by Loh and Wainwright (2012) and Lounici (2013).

**Assumption 1.** (Missing Completely at Random (MCR)) $S = \{\boldsymbol{S}_t^{(k)} \in \{0,1\}^p : t = 1, \ldots, n_k, k = 1, 2\}$ is independent of the values of $\boldsymbol{X}_t^{(1)}$ and $\boldsymbol{X}_t^{(2)}$ for $t = 1, \ldots, n_k$, $k = 1, 2$. Here $\boldsymbol{S}_t^{(k)}$ can be either deterministic or random, but independent of $\boldsymbol{X}_t^{(1)}$ and $\boldsymbol{X}_t^{(2)}$.

A major goal of the present paper is to construct a classification rule $\hat{C}$ in the high dimensional setting where $p \gg n$ for both complete and incomplete data.

## 2.2 Data-driven adaptive classifier for complete data

We first consider the case of complete data. In this setting, as mentioned in the introduction, a number of high-dimensional linear discriminant rules have been proposed in the literature. In particular, Cai and Liu (2011) introduced a classification rule called LPD (Linear Programming Discriminant) rule by directly estimating the discriminant direction

$\boldsymbol{\beta}$ through solving the following optimization problem:

$$\hat{\boldsymbol{\beta}}_{\text{LPD}} = \arg\min_{\boldsymbol{\beta}} \left\{ \|\boldsymbol{\beta}\|_1 : \text{subject to } \|\hat{\Sigma}\boldsymbol{\beta} - (\hat{\boldsymbol{\mu}}_2 - \hat{\boldsymbol{\mu}}_1)\|_\infty \leq \lambda_n \right\}, \tag{5}$$

where $\hat{\boldsymbol{\mu}}_1, \hat{\boldsymbol{\mu}}_2, \hat{\Sigma}$ are sample means and pooled sample covariance matrix respectively, and $\lambda_n = C\sqrt{\log p/n}$ is the tuning parameter with some constant $C$. Based on $\hat{\boldsymbol{\beta}}_{\text{LPD}}$, the LPD rule is then given by

$$\hat{C}_{\text{LPD}}(\boldsymbol{Z}) = \begin{cases} 1, & \hat{\boldsymbol{\beta}}_{\text{LPD}}^\top(\boldsymbol{Z} - \frac{\hat{\boldsymbol{\mu}}_1+\hat{\boldsymbol{\mu}}_2}{2}) < 0 \\ 2, & \hat{\boldsymbol{\beta}}_{\text{LPD}}^\top(\boldsymbol{Z} - \frac{\hat{\boldsymbol{\mu}}_1+\hat{\boldsymbol{\mu}}_2}{2}) \geq 0 \end{cases}. \tag{6}$$

The LPD rule is easy to implement and possesses a number of desirable properties as shown in Cai and Liu (2011). It has, however, three drawbacks. A major shortcoming of the LPD rule is that it uses a common constraint $\lambda_n$ for all coordinates of $\boldsymbol{a} = \hat{\Sigma}\boldsymbol{\beta} - (\hat{\boldsymbol{\mu}}_2 - \hat{\boldsymbol{\mu}}_1)$. This essentially treats the random vector $\boldsymbol{a}$ as homoscedastic, while in fact $\boldsymbol{a}$ is intrinsically heteroscedastic and the coordinates could have a wide range of variability. The resulting estimator $\hat{\boldsymbol{\beta}}_{\text{LPD}}$ obtained in (5) of the discriminant direction $\boldsymbol{\beta}$ has yet to be shown as rate optimal; secondly, the procedure is not adaptive in the sense that the tuning parameter $\lambda_n$ is not fully specified and needs to be chosen through an empirical method such as cross-validation. The third drawback is that the LPD rule $\hat{C}_{\text{LPD}}$ does not come with theoretical optimality guarantees.

To overcome these drawbacks, we now introduce an adaptive algorithm for the high-dimensional linear discriminant analysis with complete data, called AdaLDA (**Ada**ptive **L**inear **D**iscriminant **A**nalysis rule), which takes into account the heteroscedasticity of the random vector $\boldsymbol{a} = \hat{\Sigma}\boldsymbol{\beta} - (\hat{\boldsymbol{\mu}}_2 - \hat{\boldsymbol{\mu}}_1)$. The AdaLDA is fully data-driven and adaptive to the variability of individual entries. Before we describe the classifier in detail, it is helpful to state the following key technical result which provides the motivation for the new procedure.

**Lemma 1.** *Suppose $\{\boldsymbol{X}_t^{(1)}\}_{t=1}^{n_1}$ and $\{\boldsymbol{X}_t^{(2)}\}_{t=1}^{n_2}$ are i.i.d. random samples from $N_p(\boldsymbol{\mu}_1, \Sigma)$ and $N_p(\boldsymbol{\mu}_2, \Sigma)$ respectively with $\Sigma = (\sigma_{ij})_{1\leq i,j\leq p}$. Let $\boldsymbol{\delta} = \boldsymbol{\mu}_2 - \boldsymbol{\mu}_1$, $\boldsymbol{\beta} = \Omega\boldsymbol{\delta}$, $\Delta = \sqrt{\boldsymbol{\beta}^\top \boldsymbol{\delta}}$ and $\boldsymbol{a} = \hat{\Sigma}\boldsymbol{\beta} - (\hat{\boldsymbol{\mu}}_2 - \hat{\boldsymbol{\mu}}_1)$, where $\hat{\boldsymbol{\mu}}_1, \hat{\boldsymbol{\mu}}_2, \hat{\Sigma}$ are sample means and pooled sample covariance matrix respectively. Then with probability at least $1 - 4p^{-1}$,*

$$|a_j| \leq 4\sqrt{\sigma_{jj}} \cdot \left(\sqrt{\frac{25\Delta^2}{2} + 1}\right) \cdot \sqrt{\frac{\log p}{n}}, \quad j = 1, ..., p. \tag{7}$$

A major step in the construction of the adaptive data-driven procedure is to make the constraint in (5) adaptive to the variability of individual entries based on Lemma 1, instead of using a uniform upper bound $\lambda_n$ for all the entries. In order to apply Lemma 1, we need to estimate the diagonal elements of $\Sigma$, $\sigma_{jj}$ ($j = 1, ..., p$) and $\Delta^2$. Note that $\sigma_{jj}$ can be easily estimated by the sample variances $\hat{\sigma}_{jj}$, but $\Delta^2$ is harder to estimate.

The data-driven adaptive classification procedure AdaLDA is constructed in three steps.

**Step 1 (Estimating $\Delta^2$).** Fix $\lambda_0 = 25/2$, we estimate $\boldsymbol{\beta}$ by a preliminary estimator

$$\tilde{\boldsymbol{\beta}} = \arg\min_{\boldsymbol{\beta}} \|\boldsymbol{\beta}\|_1$$

$$\text{subject to } |\boldsymbol{e}_j^\top (\hat{\Sigma}\boldsymbol{\beta} - (\hat{\boldsymbol{\mu}}_2 - \hat{\boldsymbol{\mu}}_1))| \leq 4\sqrt{\frac{\log p}{n}} \cdot \sqrt{\hat{\sigma}_{jj}} \cdot (\lambda_0 \boldsymbol{\beta}^\top (\hat{\boldsymbol{\mu}}_2 - \hat{\boldsymbol{\mu}}_1) + 1), \; j \in [p]. \tag{8}$$

Then we estimate $\Delta^2$ by $\hat{\Delta}^2 = |\tilde{\boldsymbol{\beta}}^\top (\hat{\boldsymbol{\mu}}_2 - \hat{\boldsymbol{\mu}}_1)|$.

**Step 2 (Adaptive estimation of $\boldsymbol{\beta}$).** Given $\hat{\Delta}^2$, the final estimator $\hat{\boldsymbol{\beta}}_{\text{AdaLDA}}$ of $\boldsymbol{\beta}$ is constructed through the following linear optimization

$$\hat{\boldsymbol{\beta}}_{\text{AdaLDA}} = \arg\min_{\boldsymbol{\beta}} \|\boldsymbol{\beta}\|_1$$

$$\text{subject to } |\boldsymbol{e}_j^\top (\hat{\Sigma}\boldsymbol{\beta} - (\hat{\boldsymbol{\mu}}_2 - \hat{\boldsymbol{\mu}}_1))| \leq 4\sqrt{\frac{\log p}{n}} \cdot \sqrt{\lambda_0 \hat{\sigma}_{jj} \hat{\Delta}^2 + \hat{\sigma}_{jj}}, \; j \in [p]. \tag{9}$$

**Step 3 (Construction of AdaLDA).** The AdaLDA classification rule is obtained by plugging $\hat{\boldsymbol{\beta}}_{\text{AdaLDA}}$ into Fisher's rule (1),

$$\hat{C}_{\text{AdaLDA}}(\boldsymbol{Z}) = \begin{cases} 1, & \hat{\boldsymbol{\beta}}_{\text{AdaLDA}}^\top (\boldsymbol{Z} - \frac{\hat{\boldsymbol{\mu}}_1 + \hat{\boldsymbol{\mu}}_2}{2}) \geq 0, \\ 2, & \hat{\boldsymbol{\beta}}_{\text{AdaLDA}}^\top (\boldsymbol{Z} - \frac{\hat{\boldsymbol{\mu}}_1 + \hat{\boldsymbol{\mu}}_2}{2}) < 0. \end{cases} \tag{10}$$

This classification rule does not require a tuning parameter and the estimator $\hat{\boldsymbol{\beta}}_{\text{AdaLDA}}$ adapts to the variability of individual entries by using an entry-dependent threshold for each individual coordinate of $\hat{\Sigma}\boldsymbol{\beta} - (\hat{\boldsymbol{\mu}}_2 - \hat{\boldsymbol{\mu}}_1)$. Note that the optimization problems (8) and (9) can be cast as linear programs, so the proposed AdaLDA rule is computationally easy to implement. It will be shown in Section 3 that the AdaLDA classification rule is adaptively minimax rate optimal. Our theoretical analysis also shows that the resulting estimator $\hat{\boldsymbol{\beta}}_{\text{AdaLDA}}$ is rate optimally adaptive whenever $\lambda_0$ is a sufficiently large constant. In particular, it can be taken as fixed at $\lambda_0 = 25/2$, which is derived from the concentration inequality given in Lemma 1.

**Remark 1.** The LPD rule uses a universal tuning parameter $\lambda_n = C\sqrt{\log p/n}$ which does not take into account the heteroscedasticity of the random vector $\boldsymbol{a} = \hat{\Sigma}\boldsymbol{\beta} - (\hat{\boldsymbol{\mu}}_2 - \hat{\boldsymbol{\mu}}_1)$, and the optimality of estimation is unknown. The cross-validation method can be used to choose the tuning parameter in LPD. However, the estimator obtained through cross-validation can be variable and its theoretical properties are unclear. In contrast, the AdaLDA procedure does not depend on any unknown parameter and the estimator will be shown to be minimax rate optimal.

## 2.3 ADAM with randomly missing data

We now turn to the case of incomplete data under the MCR model. To generalize AdaLDA to the incomplete data case, we proceed by firstly estimating $\boldsymbol{\mu}_1$, $\boldsymbol{\mu}_2$ and $\Sigma$. The following estimators follow the idea in Cai and Zhang (2016), and for completeness, we present their proposed estimators below. Let

$$n_{ij}^{(k)*} = \sum_{t=1}^{n_k} S_{ti}^{(k)} S_{tj}^{(k)}, \quad 1 \leq i,j \leq p, \ k = 1, 2.$$

Here $n_{ij}^{(k)*}$ is the number of vectors $\boldsymbol{X}_t^{(k)}$ in which the $i^{th}$ and $j^{th}$ entries are both observed. In addition, we denote $n_i^{(k)*} = n_{ii}^{(k)*}$ for simplicity and

$$n_{\min}^* = \min_{i,j,k} n_{ij}^{(k)*}. \tag{11}$$

In the presence of missing values, the usual sample mean and sample covariance matrix can no longer be calculated. Instead, the "generalized sample mean" is proposed, defined by

$$\hat{\boldsymbol{\mu}}_1 = (\hat{\mu}_{1i}^*)_{1 \leq i \leq p} \quad \text{with} \quad \hat{\mu}_{1i}^* = \frac{1}{n_i^{(1)*}} \sum_{t=1}^{n_1} X_{ti}^{(1)} S_{ti}^{(1)}, \quad 1 \leq i \leq p;$$

$$\hat{\boldsymbol{\mu}}_2 = (\hat{\mu}_{2i}^*)_{1 \leq i \leq p} \quad \text{with} \quad \hat{\mu}_{2i}^* = \frac{1}{n_i^{(2)*}} \sum_{t=1}^{n_2} X_{ti}^{(2)} S_{ti}^{(2)}, \quad 1 \leq i \leq p.$$

The "generalized sample covariance matrix" is then defined by $\hat{\Sigma} = (\hat{\sigma}_{ij}^*)_{1 \leq i,j \leq p}$ with

$$\hat{\sigma}_{ij}^* = \frac{1}{n_{ij}^{(1)*} + n_{ij}^{(2)*}} \left( \sum_{t=1}^{n_1} (X_{ti}^{(1)} - \hat{\mu}_{1i}^*)(X_{tj}^{(1)} - \hat{\mu}_{1j}^*) S_{ti}^{(1)} S_{tj}^{(1)} + \sum_{t=1}^{n_2} (X_{ti}^{(2)} - \hat{\mu}_{2i}^*)(X_{tj}^{(2)} - \hat{\mu}_{2j}^*) S_{ti}^{(2)} S_{tj}^{(2)} \right).$$

For these generalized estimators, we have the following bound under the MCR model.

**Lemma 2.** *Let $\boldsymbol{\delta} = \boldsymbol{\mu}_2 - \boldsymbol{\mu}_1$, $\boldsymbol{\beta} = \Omega\boldsymbol{\delta}$, $\Delta = \sqrt{\boldsymbol{\delta}^\top \Omega \boldsymbol{\delta}}$ and $\boldsymbol{a}^* = \hat{\Sigma}\boldsymbol{\beta} - (\hat{\boldsymbol{\mu}}_2 - \hat{\boldsymbol{\mu}}_1)$. Then conditioning on $\boldsymbol{S}$, we have with high probability,*

$$|a_j^*| \leq 4\sqrt{\sigma_{jj}} \cdot \left( \sqrt{64\Delta^2 + 1} \right) \cdot \sqrt{\frac{\log p}{n_{\min}^*}}, \quad j = 1, ..., p. \tag{12}$$

**Remark 2.** Although the above result has a form that is similar to Lemma 1, its derivation is quite different and relies on a new technical tool, the large deviation bound for $\hat{\Sigma}$. This is of independent interest and is related to that of the masked sample covariance estimator considered in Levina and Vershynin (2012) and Chen et al. (2012). In particular, the masked sample covariance estimator considered in Chen et al. (2012) applies the mask matrix to the sample covariance maxtrix, while our proposed estimator $\hat{\Sigma}$ can be interpreted as applying the mask matrix to each *i.i.d.* sample, and thus is more general. The proof of Lemma 2 uses the idea of Lemma 2.1 in Cai and Zhang (2016), but yields a sharper bound. The detailed proof is given in Section A.3.2 in the supplement (Cai and Zhang, 2018).

We propose to estimate $\boldsymbol{\beta}$ adaptively and construct ADAM (**A**daptive linear **D**iscriminant **A**nalysis with randomly **M**issing data) in the following way:

**Step 1 (Estimating $\Delta^2$).** Fix $\lambda_1 = 64$. We estimate $\boldsymbol{\beta}$ by a preliminary estimator

$$\tilde{\boldsymbol{\beta}} = \arg\min_{\boldsymbol{\beta}} \|\boldsymbol{\beta}\|_1$$

$$\text{subject to } |\boldsymbol{e}_j^\top \left( \hat{\Sigma}\boldsymbol{\beta} - (\hat{\boldsymbol{\mu}}_2 - \hat{\boldsymbol{\mu}}_1) \right)| \leq 4\sqrt{\hat{\sigma}_{jj}^*} \cdot \sqrt{\frac{\log p}{n_{min}^*}} \cdot (\lambda_1 \boldsymbol{\beta}^\top(\hat{\boldsymbol{\mu}}_2 - \hat{\boldsymbol{\mu}}_1) + 1), \ j \in [p]. \tag{13}$$

Then we estimate $\Delta^2$ by $\hat{\Delta}^{*2} = |\tilde{\boldsymbol{\beta}}^\top(\hat{\boldsymbol{\mu}}_2 - \hat{\boldsymbol{\mu}}_1)|$.

**Step 2 (Adaptive estimation of $\boldsymbol{\beta}$).** Given $\hat{\Delta}^{*2}$, the final estimator $\hat{\boldsymbol{\beta}}_{\text{ADAM}}$ of $\boldsymbol{\beta}$ is constructed by the following linear optimization problem

$$\hat{\boldsymbol{\beta}}_{\text{ADAM}} = \arg\min_{\boldsymbol{\beta}} \|\boldsymbol{\beta}\|_1$$

$$\text{subject to } |\boldsymbol{e}_j^\top \left( \hat{\Sigma}\boldsymbol{\beta} - (\hat{\boldsymbol{\mu}}_2 - \hat{\boldsymbol{\mu}}_1) \right)| \leq 4\sqrt{\frac{\log p}{n_{min}^*}} \cdot \sqrt{\lambda_1 \hat{\sigma}_{jj}^* \hat{\Delta}^{*2} + \hat{\sigma}_{jj}^*}, \ j \in [p]. \tag{14}$$

**Step 3 (Construction of ADAM).** Given the estimator $\hat{\boldsymbol{\beta}}_{\text{ADAM}}$ of the discriminant direction $\boldsymbol{\beta}$, we then construct the following ADAM classification rule by plugging $\hat{\boldsymbol{\beta}}_{\text{ADAM}}$ into the oracle rule (1):

$$\hat{C}_{\text{ADAM}}(\boldsymbol{Z}) = \begin{cases} 1, & \{\boldsymbol{Z} - (\hat{\boldsymbol{\mu}}_1 + \hat{\boldsymbol{\mu}}_2)/2\}^\top \hat{\boldsymbol{\beta}}_{\text{ADAM}} \geq 0, \\ 2, & \{\boldsymbol{Z} - (\hat{\boldsymbol{\mu}}_1 + \hat{\boldsymbol{\mu}}_2)/2\}^\top \hat{\boldsymbol{\beta}}_{\text{ADAM}} < 0. \end{cases} \tag{15}$$

As shown in Section 3, $\hat{C}_{\text{ADAM}}$ has the similar theoretical performance as $\hat{C}_{\text{AdaLDA}}$.

# 3 Theoretical properties of AdaLDA and ADAM

In this section, we develop an optimality theory for high-dimensional linear discriminant analysis for both the complete data and the incomplete data settings. We first investigate the theoretical properties of the AdaLDA and ADAM algorithms proposed in Section 2 and obtain the upper bounds for the excess misclassification risk. We then establish the lower bounds for the rate of convergence. The upper and lower bounds together yield the minimax rates of convergence and show that AdaLDA and ADAM are adaptively rate optimal.

## 3.1 Theoretical Analysis of AdaLDA

We begin by considering the properties of the estimator $\hat{\boldsymbol{\beta}}_{\text{AdaLDA}}$ of the discriminant direction $\boldsymbol{\beta}$. The following theorem shows that $\hat{\boldsymbol{\beta}}_{\text{AdaLDA}}$ attains the convergence rate of $M_{n,p}\sqrt{s\log p/n}$ over the class of sparse discriminant vectors $\mathcal{G}(s, M_{n,p})$ defined in (3). The matching lower bound given in Section 3.3 implies that this rate is optimal. Therefore, AdaLDA adapts to both the sparsity pattern of $\boldsymbol{\beta}$ as well as the signal-to-noise ratio $\Delta$.

**Theorem 2.** *Consider the parameter space $\mathcal{G}(s, M_{n,p})$ with $M_{n,p} > c_L$ for some $c_L > 0$. Suppose $\boldsymbol{X}_1^{(1)}, ..., \boldsymbol{X}_{n_1}^{(1)} \overset{i.i.d.}{\sim} N_p(\boldsymbol{\mu}_1, \Sigma)$, $\boldsymbol{X}_1^{(2)}, ..., \boldsymbol{X}_{n_2}^{(2)} \overset{i.i.d.}{\sim} N_p(\boldsymbol{\mu}_2, \Sigma)$ and $n_1 \asymp n_2$. Assume that $M_{n,p}\sqrt{\frac{s\log p}{n}} = o(1)$. Then*

$$\sup_{\boldsymbol{\theta} \in \mathcal{G}(s, M_{n,p})} \mathbb{E}[\|\hat{\boldsymbol{\beta}}_{\text{AdaLDA}} - \boldsymbol{\beta}\|_2] \lesssim M_{n,p}\sqrt{\frac{s\log p}{n}}.$$

We then proceed to characterize the accuracy of the classification rule $\hat{C}_{\text{AdaLDA}}$, measured by the excess misclassification risk $R_{\boldsymbol{\theta}}(\hat{C}) - R_{\text{opt}}(\boldsymbol{\theta})$. Note that the conditional misclassification rate of $\hat{C}_{\text{AdaLDA}}$ given the two samples can be analytically calculated as

$$R_{\boldsymbol{\theta}}(\hat{C}_{\text{AdaLDA}}) = \frac{1}{2}\Phi\left(\frac{(\hat{\boldsymbol{\mu}} - \boldsymbol{\mu}_1)^\top \hat{\boldsymbol{\beta}}_{\text{AdaLDA}}}{\sqrt{\hat{\boldsymbol{\beta}}_{\text{AdaLDA}}^\top \Sigma \hat{\boldsymbol{\beta}}_{\text{AdaLDA}}}}\right) + \frac{1}{2}\bar{\Phi}\left(\frac{(\hat{\boldsymbol{\mu}} - \boldsymbol{\mu}_2)^\top \hat{\boldsymbol{\beta}}_{\text{AdaLDA}}}{\sqrt{\hat{\boldsymbol{\beta}}_{\text{AdaLDA}}^\top \Sigma \hat{\boldsymbol{\beta}}_{\text{AdaLDA}}}}\right),$$

where $\hat{\boldsymbol{\mu}} = (\hat{\boldsymbol{\mu}}_1 + \hat{\boldsymbol{\mu}}_2)/2$.

We are interested in the excess misclassification risk $R_{\boldsymbol{\theta}}(\hat{C}_{\text{AdaLDA}}) - R_{\text{opt}}(\boldsymbol{\theta})$. That is, we compare $\hat{C}_{\text{AdaLDA}}$ with the oracle Fisher's rule, whose risk is given by

$$R_{\text{opt}}(\boldsymbol{\theta}) \overset{def}{=} R_{\boldsymbol{\theta}}(C_{\boldsymbol{\theta}}) = \Phi\left(-\frac{1}{2}\Delta\right).$$

The following theorem provides an upper bound for the excess misclassification risk of the AdaLDA rule.

**Theorem 3.** *Consider the parameter space $\mathcal{G}(s, M_{n,p})$ with $M_{n,p} > c_L$ for some $c_L > 0$ and assume the conditions in Theorem 2 hold.*

1. *If $M_{n,p} \leq C_b$ for some $C_b > 0$, then there exists some constant $C > 0$,*

$$\inf_{\boldsymbol{\theta} \in \mathcal{G}(s, M_{n,p})} \mathbb{P}\left(R_{\boldsymbol{\theta}}(\hat{C}_{\text{AdaLDA}}) - R_{\text{opt}}(\boldsymbol{\theta}) \leq C \cdot \frac{s \log p}{n}\right) \geq 1 - 8p^{-1}.$$

2. *If $M_{n,p} \to \infty$ as $n \to \infty$, then there exist some constant $C > 0$ and $\delta_n = o(1)$, such that*

$$\inf_{\boldsymbol{\theta} \in \mathcal{G}(s, M_{n,p})} \mathbb{P}\left(R_{\boldsymbol{\theta}}(\hat{C}_{\text{AdaLDA}}) - R_{\text{opt}}(\boldsymbol{\theta}) \leq C \cdot e^{-(\frac{1}{8} + \delta_n)M_{n,p}^2} \cdot \frac{s \log p}{n}\right) \geq 1 - 8p^{-1}.$$

**Remark 3.** The result in Theorem 3 improves the convergence rate of the misclassification risk of the LPD rule given in Cai and Liu (2011). Consider the first case where $M_{n,p}$ is a constant not depending on $n$ and $p$, Theorem 3 of Cai and Liu (2011) shows that the convergence rate is $R_{\boldsymbol{\theta}}(\hat{C}_{\text{LPD}}) - R_{\text{opt}}(\boldsymbol{\theta}) = O_P((s \log p/n)^{1/2})$, while Theorem 3 here shows a faster rate $O_P((s \log p/n))$ when $M_{n,p}$ is a constant. The lower bounds given in Section 3.3 shows that both convergence rates in Theorem 3 are indeed optimal.

## 3.2 Theoretical Analysis of ADAM

We now investigate the theoretical properties of the ADAM procedure in the presence of missing data. Similar rates of convergence for estimation and excess misclassification risk can be obtained, but the technical analysis is much more involved under the MCR model.

Under the MCR model, suppose that the missingness pattern $S \in \{0,1\}^{n_1 \times p} \times \{0,1\}^{n_2 \times p}$ is a realization of a distribution $\mathcal{F}$. We consider the distribution space $\boldsymbol{\Psi}(n_0; n, p)$ given by

$$\boldsymbol{\Psi}(n_0; n, p) = \{\mathcal{F} : \mathbb{P}_{S \sim \mathcal{F}}\left(c_1 n_0 \leq n_{\min}^*(S) \leq c_2 n_0\right) \geq 1 - p^{-1}\},$$

for some constants $c_1, c_2 > 0$, and $n_{\min}^*(S)$ is defined for $S$ as in (11).

**Remark 4.** This distribution space includes the missing uniformly and completely at random (MUCR) model considered in Loh and Wainwright (2012); Lounici (2013) and Lounici (2014). More specifically, MUCR model assumes each entry $X_{i,j}^{(k)}$ ($k \in [2], i \in [n_k], j \in [p]$) is missing independently with probability $\epsilon$. As shown in Section A.6 in the supplement, when $\frac{1}{(1-\epsilon)^2}\sqrt{\frac{\log p}{n}} = o(1)$ as $n \to \infty$, the MUCR model is in the distribution space $\boldsymbol{\Psi}(n(1-\epsilon)^2; n, p)$.

In addition, this distribution space allows a more general variant of MUCR model that each entry $X_{i,j}^{(k)}$ is missing independently with different probabilities $\epsilon_{ij}^{(k)}$. If we assume $\tilde{c}_1 \cdot \epsilon \leq \min_{i,j,k} \epsilon_{ij}^{(k)} \leq \max_{i,j,k} \epsilon_{ij}^{(k)} \leq \tilde{c}_2 \cdot \epsilon$ for some constants $\tilde{c}_1, \tilde{c}_2 > 0$, then use the similar technique, this missingness pattern is included in $\boldsymbol{\Psi}(n(1-\epsilon)^2; n, p)$ when $\frac{1}{(1-\epsilon)^2}\sqrt{\frac{\log p}{n}} = o(1)$ as $n \to \infty$.

The following two theorems provide respectively the convergence rates for the discriminant vector estimator $\hat{\boldsymbol{\beta}}_{\text{ADAM}}$ and the excess misclassification rate of $\hat{C}_{\text{ADAM}}$ over the parameter space $\mathcal{G}(s, M_{n,p})$ for $\boldsymbol{\theta}$ and the distribution space $\boldsymbol{\Psi}(n_0; n, p)$.

**Theorem 4.** *Consider the parameter space $\mathcal{G}(s, M_{n,p})$ with $M_{n,p} > c_L$ for some $c_L > 0$ and the distribution space $\boldsymbol{\Psi}(n_0; n, p)$ with $M_{n,p}\sqrt{\frac{s \log p}{n_0}} = o(1)$. Suppose $\boldsymbol{X}_1^{(1)}, ..., \boldsymbol{X}_{n_1}^{(1)}$ and $\boldsymbol{X}_1^{(2)}, ..., \boldsymbol{X}_{n_2}^{(2)}$ are i.i.d. samples from $N_p(\boldsymbol{\mu}_1, \boldsymbol{\Sigma})$ and $N_p(\boldsymbol{\mu}_2, \boldsymbol{\Sigma})$ respectively. Assuming that $\boldsymbol{X}_1^{*(1)}, ..., \boldsymbol{X}_{n_1}^{*(1)}$ and $\boldsymbol{X}_1^{*(2)}, ..., \boldsymbol{X}_{n_2}^{*(2)}$ defined in (4) is observed and Assumption 1 with $S = \{\boldsymbol{S}_t^{(k)}\}_{t \in [n_k], k \in [2]}$ holds. Then the risk of estimating the discriminant direction $\boldsymbol{\beta}$ by ADAM satisfies*

$$\sup_{\substack{\boldsymbol{\theta} \in \mathcal{G}(s, M_{n,p}) \\ \mathcal{F} \in \boldsymbol{\Psi}(n_0; n, p)}} \mathbb{E}[\|\hat{\boldsymbol{\beta}}_{\text{ADAM}} - \boldsymbol{\beta}\|_2] \lesssim M_{n,p}\sqrt{\frac{s \log p}{n_0}}.$$

**Theorem 5.** *Suppose the conditions of Theorem 4 hold.*

1. *If $M_{n,p} \leq C_b$ for some $C_b > 0$, then there exists some constant $C > 0$, such that*

$$\inf_{\substack{\boldsymbol{\theta} \in \mathcal{G}(s, M_{n,p}) \\ \mathcal{F} \in \boldsymbol{\Psi}(n_0; n, p)}} \mathbb{P}\left(R_{\boldsymbol{\theta}}(\hat{C}_{\text{ADAM}}) - R_{\text{opt}}(\boldsymbol{\theta}) \leq C \cdot \frac{s \log p}{n_0}\right) \geq 1 - 12p^{-1}.$$

2. *If $M_{n,p} \to \infty$ as $n \to \infty$, then there exist some constant $C > 0$ and $\delta_n = o(1)$, such that*

$$\inf_{\substack{\boldsymbol{\theta} \in \mathcal{G}(s, M_{n,p}) \\ \mathcal{F} \in \boldsymbol{\Psi}(n_0; n, p)}} \mathbb{P}\left(R_{\boldsymbol{\theta}}(\hat{C}_{\text{ADAM}}) - R_{\text{opt}}(\boldsymbol{\theta}) \leq C \cdot e^{-(\frac{1}{8}+\delta_n)M_{n,p}^2} \cdot \frac{s \log p}{n_0}\right) \geq 1 - 12p^{-1}.$$

In the complete data case, we have $n_0 = n$, so the rates of convergence shown in Theorem 4 and 5 match those in Theorems 2 and 3. In addition, in the special case of MUCR model, Theorem 4 and 5 imply the following result.

**Corollary 1.** *Under the conditions of Theorem 3 and consider the MUCR model with missing probability $\epsilon$. If $(M_{n,p}^2 \frac{s \log p}{n} \vee \sqrt{\frac{\log p}{n}}) \cdot \frac{1}{(1-\epsilon)^2} = o(1)$, then the risk of estimating the discriminant direction $\boldsymbol{\beta}$ by ADAM over the class $G(s, M_{n,p})$ satisfies*

$$\sup_{\boldsymbol{\theta} \in \mathcal{G}(s, M_{n,p})} \mathbb{E}[\|\hat{\boldsymbol{\beta}}_{\text{ADAM}} - \boldsymbol{\beta}\|_2] \lesssim M_{n,p}\sqrt{\frac{s \log p}{n(1-\epsilon)^2}}.$$

*Moreover, there exist constant $C > 0$ and $\delta_n = o(1)$, such that the excess misclassification risk over the class $G(s, M_{n,p})$ satisfies*

$$\inf_{\boldsymbol{\theta} \in \mathcal{G}(s, M_{n,p})} \mathbb{P}\left(R_{\boldsymbol{\theta}}(\hat{C}_{\text{ADAM}}) - R_{\text{opt}}(\boldsymbol{\theta}) \leq C \cdot e^{-(\frac{1}{8}+\delta_n)M_{n,p}^2} \cdot \frac{s \log p}{n(1-\epsilon)^2}\right) \geq 1 - 13p^{-1}.$$

This result shows that, although the sample size only loses a proportion of $\epsilon$, the convergence rates for the estimation risk and misclassification rate shrunk at the rate of $n(1-\epsilon)^2$ under the MUCR model.

## 3.3 Minimax lower bounds

To understand the difficulty of the classification problem and the related estimation problem as well as to establish the optimality for the AdaLDA and ADAM classifiers, it is essential to obtain the minimax lower bounds for the estimation risk and the excess misclassification risk. In this section, we only state the results for the missing data setting as the complete data setting can be treated as a special case. The following lower bound results show that the rates of convergence obtained by AdaLDA and ADAM algorithms are indeed optimal, for both estimation of the discriminant direction $\boldsymbol{\beta}$ and classification.

**Theorem 6.** *Consider the parameter space $\mathcal{G}(s, M_{n,p})$ with $M_{n,p} > c_L$ for some $c_L > 0$ and the distribution space $\boldsymbol{\Psi}(n_0; n, p)$ with $M_{n,p}\sqrt{\frac{s \log p}{n_0}} = o(1)$. For any $n_0 > 1$, suppose $1 \leq s \leq o(\frac{n_0}{\log p})$ and $\frac{\log p}{\log(p/s)} = O(1)$. Then under MCR model, the minimax risk of estimating the discriminant direction $\boldsymbol{\beta}$ over the class $G(s, M_{n,p})$ and $\boldsymbol{\Psi}(n_0; n, p)$ satisfies*

$$\inf_{\hat{\boldsymbol{\beta}}} \sup_{\substack{\boldsymbol{\theta} \in \mathcal{G}(s, M_{n,p}) \\ \mathcal{F} \in \boldsymbol{\Psi}(n_0; n, p)}} \mathbb{E}[\|\hat{\boldsymbol{\beta}} - \boldsymbol{\beta}\|_2] \gtrsim M_{n,p}\sqrt{\frac{s \log p}{n_0}}.$$

**Theorem 7.** *Consider the parameter space $\mathcal{G}(s, M_{n,p})$ with $M_{n,p} > c_L$ for some $c_L > 0$ and the distribution space $\boldsymbol{\Psi}(n_0; n, p)$ with $M_{n,p}\sqrt{\frac{s \log p}{n_0}} = o(1)$. For any $n_0 \geq 1$, suppose $1 \leq s \leq o(\frac{n_0}{\log p})$ and $\frac{\log p}{\log(p/s)} = O(1)$. Then under the MCR model, the minimax risk of the excess misclassification error over the class $G(s, M_{n,p})$ and $\boldsymbol{\Psi}(n_0; n, p)$ satisfies that*

1. *If $M_{n,p} \leq C_b$ for some $C_b > 0$, then for any $\alpha > 0$, there are some constants $C_\alpha > 0$ such that*

$$\inf_{\hat{C}} \sup_{\substack{\boldsymbol{\theta} \in \mathcal{G}(s, M_{n,p}) \\ \mathcal{F} \in \boldsymbol{\Psi}(n_0; n, p)}} \mathbb{P}(R_{\boldsymbol{\theta}}(\hat{C}) - R_{\text{opt}}(\boldsymbol{\theta}) \geq C_\alpha \cdot \frac{s \log p}{n_0}) \geq 1 - \alpha.$$

2. *If $M_{n,p} \to \infty$ as $n \to \infty$, then for any $\alpha > 0$, there are some constants $C_\alpha > 0$ and $\tilde{\delta}_n = o(1)$ such that*

$$\inf_{\hat{C}} \sup_{\substack{\boldsymbol{\theta} \in \mathcal{G}(s, M_{n,p}) \\ \mathcal{F} \in \boldsymbol{\Psi}(n_0; n, p)}} \mathbb{P}(R_{\boldsymbol{\theta}}(\hat{C}) - R_{\text{opt}}(\boldsymbol{\theta}) \geq C_\alpha \cdot e^{-(\frac{1}{8} + \tilde{\delta}_n) M_{n,p}^2} \cdot \frac{s \log p}{n_0}) \geq 1 - \alpha.$$

**Remark 5.** In the complete data case, $n_{min}^* = \min\{n_1, n_2\} = n$, so Theorems 6 and 7 together with Theorems 1-4 imply that both AdaLDA and ADAM algorithms attain the optimal rates of convergence in terms of estimation and classification error.

We should also note that the proof of Theorem 7 is not straightforward. This is partially due to the fact that the excess risk $R_{\boldsymbol{\theta}}(\hat{C}) - R_{\text{opt}}(\boldsymbol{\theta})$ does not satisfy the triangle inequality that is required by standard lower bound techniques. A key technique here is to make a connection to an alternative risk function. For a generic classification rule $\hat{C}$, we define

$$L_{\boldsymbol{\theta}}(\hat{C}) = \mathbb{P}_{\boldsymbol{\theta}}(\hat{C}(\boldsymbol{Z}) \neq C_{\boldsymbol{\theta}}(\boldsymbol{Z})), \tag{16}$$

where $C_{\boldsymbol{\theta}}(\boldsymbol{Z})$ is the Fisher's linear discriminant rule in (1). The following lemma enables us to reduce the loss $R_{\boldsymbol{\theta}}(\hat{C}) - R_{\text{opt}}(\boldsymbol{\theta})$ to the risk function $L_{\boldsymbol{\theta}}(\hat{C})$.

**Lemma 3.** *Let* $\boldsymbol{Z} \sim \frac{1}{2} N_p(\boldsymbol{\mu}_1, \Sigma) + \frac{1}{2} N_p(\boldsymbol{\mu}_2, \Sigma)$ *with parameter* $\boldsymbol{\theta} = (\boldsymbol{\mu}_1, \boldsymbol{\mu}_2, \Sigma)$. *If a classifier* $\hat{C}$ *satisfying* $L_{\boldsymbol{\theta}}(\hat{C}) = o(1)$ *as* $n \to \infty$, *then for sufficiently large* $n$,

$$R_{\boldsymbol{\theta}}(\hat{C}) - R_{\text{opt}}(\boldsymbol{\theta}) \geq \frac{\sqrt{2\pi}\Delta}{8} e^{\Delta^2/8} \cdot L_{\boldsymbol{\theta}}^2(\hat{C}).$$

Lemma 3 shows the relationship between the risk function $R_{\boldsymbol{\theta}}(\hat{C}) - R_{\text{opt}}(\boldsymbol{\theta})$ and a more "standard" risk function $L_{\boldsymbol{\theta}}(\hat{C})$, who has the following property which served the same purpose as the triangle inequality.

**Lemma 4.** *Let* $\boldsymbol{\theta} = (\boldsymbol{\mu}, -\boldsymbol{\mu}, I_p)$ *and* $\tilde{\boldsymbol{\theta}} = (\tilde{\boldsymbol{\mu}}, -\tilde{\boldsymbol{\mu}}, I_p)$ *with* $\|\boldsymbol{\mu}\|_2 = \|\tilde{\boldsymbol{\mu}}\|_2 = \Delta/2$. *For any classifier* $C$, *if* $\|\boldsymbol{\mu} - \tilde{\boldsymbol{\mu}}\|_2 = o(1)$ *as* $n \to \infty$, *then for sufficiently large* $n$,

$$L_{\boldsymbol{\theta}}(C) + L_{\tilde{\boldsymbol{\theta}}}(C) \geq \frac{1}{\Delta} e^{-\Delta^2/8} \cdot \|\boldsymbol{\mu} - \tilde{\boldsymbol{\mu}}\|_2.$$

Using Lemmas 3 and 4, we can then use Fano's inequality to complete the proof of Theorem 7. The details are shown in Section 5.

In addition, similar minimax lower bounds for estimating $\boldsymbol{\beta}$ and the excess misclassification error can be established under the MUCR model. The following result shows that the convergence rates in Corollary 1 are minimax rate optimal.

**Theorem 8.** *Under the conditions of Theorem 6 and MUCR model with missing probability* $\epsilon$, *and further assume that* $((M_{n,p}^2 \frac{s \log p}{n}) \vee \sqrt{\frac{\log p}{n}}) \cdot \frac{1}{(1-\epsilon)^2} = o(1)$, *then the minimax risk of estimating the discriminant direction* $\boldsymbol{\beta}$ *by ADAM over the class* $G(s, M_{n,p})$ *under the MUCR model satisfies*

$$\inf_{\hat{\boldsymbol{\beta}}} \sup_{\boldsymbol{\theta} \in G(s, M_{n,p})} \mathbb{E}[\|\hat{\boldsymbol{\beta}} - \boldsymbol{\beta}\|_2] \gtrsim M_{n,p} \sqrt{\frac{s \log p}{n(1-\epsilon)^2}}.$$

*Moreover, if* $M_{n,p} \to \infty$ *and* $\epsilon < 1 - c_B$ *for some* $c_B \in (0, 1)$, *the minimax risk of the misclassification error over the class* $G(s, M_{n,p})$ *satisfies that for any* $\alpha, \delta > 0$, *there are some constants* $C_\alpha > 0$, *such that*

$$\inf_{\hat{C}} \sup_{\boldsymbol{\theta} \in \mathcal{G}(s, M_{n,p})} \mathbb{P}(R_{\boldsymbol{\theta}}(\hat{C}) - R_{\text{opt}}(\boldsymbol{\theta}) \geq C_\alpha \cdot e^{-(\frac{1}{8}+\delta)M_{n,p}^2} \cdot \frac{s \log p}{n(1-\epsilon)^2}) \geq 1 - \alpha.$$

# 4 Numerical results

The proposed AdaLDA and ADAM classifiers are easy to implement, and the `MATLAB` code is available at `https://github.com/linjunz/ADAM`. We investigate in this section the numerical performance of AdaLDA and ADAM using both simulated and real data.

## 4.1 Simulations

In all the simulations, the sample size is $n_1 = n_2 = 200$ while the number of variables $p$ varies from $100, 200$ to $400$. The probability of being in either of the two classes is equal. The discriminant vector $\boldsymbol{\beta} = (1, \ldots, 1, 0, \ldots, 0)^\top$ is sparse such that only the first $s = 10$ entries are nonzero. We consider the following three models for the covariance matrix $\Sigma$.

**Model 1 Erdős-Rényi random graph:** Let $\tilde{\Omega} = (\tilde{\omega}_{ij})$ where $\tilde{\omega}_{ij} = u_{ij}\delta_{ij}$, $\delta_{ij} \sim$ Ber$(\rho)$ being the Bernoulli random variable with success probability $0.2$ and $u_{ij} \sim$ Unif$[0.5, 1] \cup [-1, -0.5]$. After symmetrizing $\tilde{\Omega}$, set $\Omega = \tilde{\Omega} + \{\max(-\phi_{\min}(\tilde{\Omega}), 0) + 0.05\}I_p$ to ensure the positive definiteness. Finally, $\Omega$ is standardized to have unit diagonals and $\Sigma = \Omega^{-1}$.

**Model 2 Block sparse model:** $\Omega = (\mathbf{B} + \delta I_p)/(1+\delta)$ where $b_{ij} = b_{ji} = 0.5 \times $ Ber$(1, 0.2)$ for $1 \leq i \leq p/2$, $i < j \leq p$; $b_{ij} = b_{ji} = 0.5$ for $p/2 + 1 \leq i < j \leq p$; $b_{ii} = 1$ for $1 \leq i \leq p$. In other words, only the first $p/2$ rows and columns of $\Omega$ are sparse, whereas the rest of the matrix is not sparse. Here $\delta = \max(-\phi_{\min}(\mathbf{B}), 0) + 0.05$. The matrix $\Omega$ is also standardized to have unit diagonals and $\Sigma = \Omega^{-1}$.

**Model 3 AR(1) model:** $(\Sigma_{ij})_{p \times p}$ with $\Sigma_{ij} = 0.9^{|i-j|}$.

Given the covariance matrix $\Sigma$ generated by the model above, the means are $\boldsymbol{\mu}_1 = (0, \ldots, 0)^\top$ and $\boldsymbol{\mu}_2 = \boldsymbol{\mu}_1 - \Sigma\boldsymbol{\beta}$. The missing mechanism is chosen such that each entry $X_{ki}$ is observed with probability $p = 1 - \epsilon \in (0, 1)$. We change the missing proportion $\epsilon$ from 0 to 0.2. We apply AdaLDA rule when the data is complete, i.e. $\epsilon = 0$, and apply ADAM rule when $\epsilon > 0$. The AdaLDA rule is then compared with the LPD (Cai and Liu, 2011), SLDA (Shao et al., 2011), FAIR (Fan and Fan, 2008), and NSC (Tibshirani et al., 2002) rules whose tuning parameters are chosen by cross-validation. We also note that one commonly used method, the Naive Bayes rule is a special case of the NSC rule with tuning parameter $\lambda_\Delta = 0$, so it's not included in the comparison. The misclassification errors are recorded in the following table. For each setting, the number of repetition is set to 100.

It can be seen from the simulation results that the proposed AdaLDA algorithm, which is purely data-driven and tuning free, has a very similar performance to that of the LPD algorithm with optimally chosen tuning parameters and outperforms all the other methods. In addition, ADAM does not lose much accuracy in the presence of missing data.

Table 1: Misclassification errors (%) for Model 1

| Method | ADAM | | | | AdaLDA | LPD | SLDA | FAIR | NSC | Oracle |
|---|---|---|---|---|---|---|---|---|---|---|
| $(s,p)\backslash \epsilon$ | 0.2 | 0.15 | 0.1 | 0.05 | 0 | 0 | 0 | 0 | 0 | 0 |
| (10,100) | 16.82 | 15.97 | 15.10 | 14.33 | 12.89 | 12.23 | 18.42 | 16.34 | 17.43 | 10.22 |
| (20,100) | 12.63 | 12.17 | 11.92 | 11.68 | 11.61 | 12.08 | 28.76 | 15.02 | 15.60 | 7.06 |
| (10,200) | 29.43 | 28.11 | 27.94 | 2658 | 25.90 | 27.32 | 39.87 | 33.22 | 37.78 | 21.50 |
| (20,200) | 18.89 | 17.78 | 17.72 | 17.20 | 15.42 | 14.67 | 26.78 | 18.87 | 21.13 | 11.63 |
| (10,400) | 31.34 | 30.28 | 29.40 | 29.09 | 28.92 | 30.89 | 37.33 | 46.45 | 36.12 | 25.46 |
| (20,400) | 26.22 | 25.78 | 24.21 | 23.54 | 22.07 | 23.96 | 34.78 | 32.45 | 33.82 | 15.63 |

Table 2: Misclassification errors (%) for Model 2

| Method | ADAM | | | | AdaLDA | LPD | SLDA | FAIR | NSC | Oracle |
|---|---|---|---|---|---|---|---|---|---|---|
| $(s,p)\backslash \epsilon$ | 0.2 | 0.15 | 0.1 | 0.05 | 0 | 0 | 0 | 0 | 0 | 0 |
| (10,100) | 3.69 | 3.59 | 3.42 | 3.38 | 3.31 | 3.40 | 3.48 | 4.03 | 4.41 | 3.00 |
| (20,100) | 0.33 | 0.30 | 0.29 | 0.25 | 0.20 | 0.22 | 0.24 | 0.24 | 0.37 | 0.14 |
| (10,200) | 5.07 | 4.73 | 4.13 | 4.03 | 3.80 | 3.73 | 3.98 | 4.95 | 5.59 | 3.39 |
| (20,200) | 0.60 | 0.53 | 0.43 | 0.42 | 0.40 | 0.46 | 0.53 | 0.82 | 0.67 | 0.33 |
| (10,400) | 5.54 | 5.18 | 5.05 | 4.21 | 4.04 | 4.11 | 4.20 | 5.20 | 6.89 | 3.68 |
| (20,400) | 1.12 | 1.00 | 0.76 | 0.67 | 0.61 | 0.65 | 0.81 | 1.38 | 1.30 | 0.41 |

Table 3: Misclassification errors (%) for Model 3

| Method | ADAM | | | | AdaLDA | LPD | SLDA | FAIR | NSC | Oracle |
|---|---|---|---|---|---|---|---|---|---|---|
| $(s,p)\backslash \epsilon$ | 0.2 | 0.15 | 0.1 | 0.05 | 0 | 0 | 0 | 0 | 0 | 0 |
| (10,100) | 18.91 | 18.72 | 18.58 | 18.16 | 18.04 | 17.99 | 23.08 | 23.72 | 23.00 | 11.78 |
| (20,100) | 19.93 | 18.98 | 18.92 | 18.13 | 17.63 | 17.69 | 23.01 | 23.44 | 22.65 | 10.78 |
| (10,200) | 19.15 | 18.38 | 18.00 | 17.84 | 17.41 | 17.94 | 22.67 | 23.87 | 23.71 | 11.78 |
| (20,200) | 18.50 | 18.36 | 18.20 | 17.81 | 17.66 | 17.77 | 23.00 | 24.36 | 24.04 | 10.74 |
| (10,400) | 18.95 | 18.73 | 18.59 | 18.27 | 17.86 | 18.66 | 24.93 | 27.29 | 26.08 | 11.71 |
| (20,400) | 18.47 | 18.39 | 18.25 | 18.10 | 17.82 | 18.14 | 24.78 | 26.77 | 26.49 | 10.68 |

## 4.2 Real data analysis

In addition to the simulation studies, we also illustrate the merits of the AdaLDA and ADAM classifiers in an analysis of two real datasets to further investigate the numerical performance of the proposed methods. One dataset, available at www.chestsurg.org, is the Lung cancer data analyzed by Gordon et al. (2002). Another dataset is the Leukemia data from high-density Affymetrix oligonucleotide arrays that was previously analyzed in Golub et al. (1999), and is available at www.broad.mit.edu/cgi-bin/cancer/datasets.cgi. These two datasets were frequently used for illustrating the empirical performance of the classifier for high-dimensional data in recent literature. We will compare AdaLDA and ADAM with the existing methods.

### 4.2.1 Lung cancer data

We evaluate the proposed methods by classifying between malignant pleural mesothelioma (MPM) and adenocarcinoma (ADCA) of the lung. There are 181 tissue samples (31 MPM and 150 ADCA) and each sample is described by 12533 genes in the lung cancer dataset in Gordon et al. (2002). This dataset has been analyzed in Fan and Fan (2008) using FAIR and NSC. In this section we apply the AdaLDA and ADAM rules to this dataset for disease classification. When ADAM rule is used, we make each entry in the dataset missing uniformly and independently with probability $\epsilon$. In the simulation, given the small sample size, we choose $\epsilon = 0.05$ and $\epsilon = 0.1$.

The sample variances of the genes range over a wide interval. We first compute the sample variances for each gene and drop the lower and upper 6-quantiles to control the condition number of $\hat{\Sigma}$. The average misclassification errors are computed by using 5-fold cross-validation for various methods with 50 repetitions. To reduce the computational costs, in each repetition, only 1500 genes with the largest absolute values of the two sample $t$ statistics are used. As seen in the Table 4, the classification result of AdaLDA is better than existing methods, including LPD (Cai and Liu, 2011), FAIR (Fan and Fan, 2008), and NSC (Tibshirani et al., 2002) methods, although only 1500 genes were used. Moreover, in the incomplete data case, ADAM still has satisfactory accuracy.

Table 4: Classification error of Lung cancer data by various methods

|  | ADAM$_{(\epsilon=0.1)}$ | ADAM$_{(\epsilon=0.05)}$ | AdaLDA | LPD | SLDA | FAIR | NSC |
|---|---|---|---|---|---|---|---|
| Testing error | 5.53% | 3.22% | 2.09% | 2.11% | 4.88% | 3.64% | 7.30% |

Table 5: Classification error of Leukemia data by various methods

|  | ADAM$_{(\epsilon=0.1)}$ | ADAM$_{(\epsilon=0.05)}$ | AdaLDA | LPD | FAIR | SLDA | NSC |
|---|---|---|---|---|---|---|---|
| Testing error | 8.47% | 7.53% | 2.94% | 3.09% | 2.94% | 5.76% | 8.82% |

#### 4.2.2 Leukemia data

Golub et al. (1999) applied gene expression microarray techniques to study human acute leukemia and discovered the distinction between acute myeloid leukemia (AML) and acute lymphoblastic leukemia (ALL). There are 72 tissue samples (47 ALL and 25 AML) and 7129 genes in the Leukemia dataset. In this section, we apply the AdaLDA rule to this dataset and compare the classification results with those obtained by LPD (Cai and Liu, 2011), FAIR, NSC (Fan and Fan, 2008), SLDA (Shao et al., 2011) and Naive Bayes (NB) methods. Same as the analysis of lung cancer data, when ADAM rule is used, we make each entry in the dataset missing independently with probability $\epsilon \in \{0.05, 0.1\}$

As in the analysis of the lung cancer data, we first drop genes with extreme sample variances out of lower and upper 6-quantiles. Similar to the analysis of the lung cancer data, the average misclassification errors are computed by using two-fold cross-validation for various methods with 50 repetitions, and to control the computational costs, we use 2000 genes with the largest absolute values of the two sample $t$ statistics in each repetition. Classification results are summarized in Table 5. The AdaLDA has the similar performance as the LPD rule and FAIR, as obtain the misclassification error of about 3%. In contrast, the navie-Bayes rule misclassifies 20.59% testing samples and SLDA misclassifies 5.76% testing samples. Fan and Fan (2008) report a test error rate of 2.94% for FAIR and a test error rate of 8.82% for NSC proposed by Tibshirani et al. (2002). In the presence of missing data, ADAM misclassifies 7.53% and 8.47% testing samples when the missing proportion is 0.05 and 0.1 respectively.

## 5 Proofs

In this section, we prove the main results, Theorem 2, 3, 4 5, 6 and 7. Theorem 1 follows from Theorems 3 and 7. Since $n_1 \asymp n_2$, without loss of the generality we shall assume $n_1 = n_2 = n$ in the proofs. For reasons of space, the proofs of the technical lemmas are given in the Supplementary Material (Cai and Zhang, 2018).

## 5.1 Proof of Theorem 2

To prove Theorem 2 we begin by collecting a few important technical lemmas that will be used in the main proofs.

### 5.1.1 Auxiliary Lemmas

**Lemma 5.** *Suppose $\mathbf{X}_1, ..., \mathbf{X}_n$ i.i.d. $\sim N_p(\boldsymbol{\mu}, \Sigma)$, and assume that $\hat{\boldsymbol{\mu}}$, $\hat{\Sigma}$ are the sample mean and sample covariance matrix respectively. Let $\Gamma(s) = \{\boldsymbol{u} \in \mathbb{R}^p : \|\boldsymbol{u}_{S^C}\|_1 \leq \|\boldsymbol{u}_S\|_1$, for some $S \subset [p]$ with $|S| = s\}$, then with probability at least $1 - p^{-1}$,*

$$\sup_{\boldsymbol{u} \in \Gamma(s)} \boldsymbol{u}^\top (\hat{\boldsymbol{\mu}} - \boldsymbol{\mu}) \lesssim \sqrt{\frac{s \log p}{n}};$$

$$\sup_{\boldsymbol{u}, \boldsymbol{v} \in \Gamma(s)} \boldsymbol{u}^\top (\hat{\Sigma} - \Sigma) \boldsymbol{v} \lesssim \sqrt{\frac{s \log p}{n}}.$$

**Lemma 6.** *Suppose $\boldsymbol{x}, \boldsymbol{y} \in \mathbb{R}^p$. Let $\boldsymbol{h} = \boldsymbol{x} - \boldsymbol{y}$ and $\mathcal{S} = \mathrm{supp}(\boldsymbol{y})$. If $\|\boldsymbol{x}\|_1 \leq \|\boldsymbol{y}\|_1$, then $\boldsymbol{h} \in \Gamma(s)$ with $s = |S|$, that is,*

$$\|\boldsymbol{h}_{\mathcal{S}^c}\|_1 \leq \|\boldsymbol{h}_S\|_1.$$

### 5.1.2 Main proof of Theorem 2

Recall that $\hat{\boldsymbol{\beta}}_{\mathrm{AdaLDA}}$ is constructed by the following two steps.

**Step 1.** Estimating $\Delta^2$

$$\tilde{\boldsymbol{\beta}} = \arg\min_{\boldsymbol{\beta}} \left\{ |\boldsymbol{e}_j^\top \left( \hat{\Sigma}\boldsymbol{\beta} - (\hat{\boldsymbol{\mu}}_2 - \hat{\boldsymbol{\mu}}_1) \right) | \leq 4\sqrt{\frac{\log p}{n}} \cdot \sqrt{\hat{\sigma}_{jj}} \cdot (\lambda_0 \boldsymbol{\beta}^\top (\hat{\boldsymbol{\mu}}_2 - \hat{\boldsymbol{\mu}}_1) + 1), \, j \in [p] \right\}. \tag{17}$$

Then we estimate $\Delta^2$ by $\hat{\Delta}^2 = |\langle \tilde{\boldsymbol{\beta}}, \hat{\boldsymbol{\mu}}_2 - \hat{\boldsymbol{\mu}}_1 \rangle|$.

**Step 2.** Adaptive estimation of $\boldsymbol{\beta}$. Given $\hat{\Delta}^2$, the final estimator $\hat{\boldsymbol{\beta}}_{\mathrm{AdaLDA}}$ of $\boldsymbol{\beta}$ is constructed by the following linear optimization problem

$$\hat{\boldsymbol{\beta}}_{\mathrm{AdaLDA}} = \arg\min_{\boldsymbol{\beta}} \left\{ |\boldsymbol{e}_j^\top \left( \hat{\Sigma}\boldsymbol{\beta} - (\hat{\boldsymbol{\mu}}_2 - \hat{\boldsymbol{\mu}}_1) \right) | \leq 4\sqrt{\frac{\log p}{n}} \cdot \sqrt{\lambda_0 \hat{\sigma}_{jj} \hat{\Delta}^2 + \hat{\sigma}_{jj}}, \, j \in [p] \right\}. \tag{18}$$

Firstly, let's show the consistency of estimating $\Delta^2$. Recall the definition of $\tilde{\boldsymbol{\beta}}$ and using

Lemma 5, we have with high probability at least $1 - 3p^{-1}$,

$$\begin{aligned}|(\tilde{\boldsymbol{\beta}} - \boldsymbol{\beta})^\top \Sigma (\tilde{\boldsymbol{\beta}} - \boldsymbol{\beta})| \leq &|(\tilde{\boldsymbol{\beta}} - \boldsymbol{\beta})^\top (\hat{\Sigma}\tilde{\boldsymbol{\beta}} - \hat{\boldsymbol{\delta}})| + |(\tilde{\boldsymbol{\beta}} - \boldsymbol{\beta})^\top (\hat{\Sigma} - \Sigma)\tilde{\boldsymbol{\beta}}| + |(\tilde{\boldsymbol{\beta}} - \boldsymbol{\beta})^\top (\boldsymbol{\delta} - \hat{\boldsymbol{\delta}})| \\ \leq &\|\tilde{\boldsymbol{\beta}} - \boldsymbol{\beta}\|_1 \|\hat{\Sigma}\tilde{\boldsymbol{\beta}} - \hat{\boldsymbol{\delta}}\|_\infty + |(\tilde{\boldsymbol{\beta}} - \boldsymbol{\beta})^\top (\hat{\Sigma} - \Sigma)(\tilde{\boldsymbol{\beta}} - \boldsymbol{\beta}))| + |(\tilde{\boldsymbol{\beta}} - \boldsymbol{\beta})^\top (\hat{\Sigma} - \Sigma)\boldsymbol{\beta})| \\ &+ |(\tilde{\boldsymbol{\beta}} - \boldsymbol{\beta})^\top (\boldsymbol{\delta} - \hat{\boldsymbol{\delta}})| \\ \lesssim & \sqrt{s}\|\tilde{\boldsymbol{\beta}} - \boldsymbol{\beta}\|_2 \cdot \|\hat{\Sigma}\tilde{\boldsymbol{\beta}} - \hat{\boldsymbol{\delta}}\|_\infty + \|\tilde{\boldsymbol{\beta}} - \boldsymbol{\beta}\|_2 \cdot \sqrt{\frac{s \log p}{n}} \cdot \|\boldsymbol{\beta} - \tilde{\boldsymbol{\beta}}\|_2 \\ &+ \|\boldsymbol{\beta} - \tilde{\boldsymbol{\beta}}\|_2 \sqrt{\frac{s \log p}{n}} \cdot \|\boldsymbol{\beta}\|_2 + \|\boldsymbol{\beta} - \tilde{\boldsymbol{\beta}}\|_2 \sqrt{\frac{s \log p}{n}},\end{aligned} \tag{19}$$

where the third inequality uses Lemma 5 and the fact that $\boldsymbol{\beta}, \tilde{\boldsymbol{\beta}} - \boldsymbol{\beta} \in \Gamma(s)$. In fact, $\boldsymbol{\beta}$ is a feasible solution to (8) due to Lemma 1 and thus $\|\tilde{\boldsymbol{\beta}}\|_1 \leq \|\boldsymbol{\beta}\|_1$. Then by Lemma 6, we have $\tilde{\boldsymbol{\beta}} - \boldsymbol{\beta} \in \Gamma(s)$. In addition, $\|\boldsymbol{\beta}\|_0 \leq s$, so we have $\boldsymbol{\beta} \in \Gamma(s)$.

In addition, by standard derivation of the accuracy of sample variance, since $M^{-1} \leq \lambda_{\min}(\Sigma) \leq \lambda_{\max}(\Sigma) \leq M$, by using the union bound technique, we have with probability at least $1 - p^{-1}$,

$$\max_{i \in [p]} |\hat{\sigma}_{ii} - \sigma_{ii}| \lesssim \sqrt{\frac{\log p}{n}},$$

which implies with probability at least $1 - p^{-1}$,

$$\max_{i \in [p]} |\hat{\sigma}_{ii}| \leq 2M.$$

In addition, since $\Delta \geq M_{n,p} \geq c_L > 0$, then with probability at least $1 - 3p^{-1}$,

$$\begin{aligned}\|\hat{\Sigma}\tilde{\boldsymbol{\beta}} - \hat{\boldsymbol{\delta}}\|_\infty \leq & 4\sqrt{\frac{\log p}{n}} \cdot \sqrt{\hat{\sigma}_{jj}} \cdot (\lambda_0 \tilde{\boldsymbol{\beta}}^\top (\hat{\boldsymbol{\mu}}_2 - \hat{\boldsymbol{\mu}}_1) + 1) \\ \lesssim & \sqrt{\frac{\log p}{n}} \cdot |(\tilde{\boldsymbol{\beta}} - \boldsymbol{\beta})^\top (\hat{\boldsymbol{\mu}}_2 - \hat{\boldsymbol{\mu}}_1) + 1| + \sqrt{\frac{\log p}{n}} \cdot |\boldsymbol{\beta}^\top (\hat{\boldsymbol{\mu}}_2 - \hat{\boldsymbol{\mu}}_1)| \\ \leq & \sqrt{\frac{\log p}{n}} \cdot (|(\tilde{\boldsymbol{\beta}} - \boldsymbol{\beta})^\top (\boldsymbol{\mu}_2 - \boldsymbol{\mu}_1)| + |(\tilde{\boldsymbol{\beta}} - \boldsymbol{\beta})^\top (\hat{\boldsymbol{\mu}}_2 - \hat{\boldsymbol{\mu}}_1 - \boldsymbol{\mu}_2 + \boldsymbol{\mu}_1)| + 1) \\ &+ \sqrt{\frac{\log p}{n}} \cdot (|\boldsymbol{\beta}^\top (\boldsymbol{\mu}_2 - \boldsymbol{\mu}_1)| + |\boldsymbol{\beta}^\top (\boldsymbol{\mu}_2 - \boldsymbol{\mu}_1 - \hat{\boldsymbol{\mu}}_2 + \hat{\boldsymbol{\mu}}_1)|) \\ \lesssim & \sqrt{\frac{\log p}{n}} \Delta \|\tilde{\boldsymbol{\beta}} - \boldsymbol{\beta}\|_2 + \sqrt{s} \cdot \frac{\log p}{n} \|\tilde{\boldsymbol{\beta}} - \boldsymbol{\beta}\|_2 + \sqrt{\frac{\log p}{n}} \Delta^2 + \sqrt{s} \cdot \frac{\log p}{n} \Delta,\end{aligned}$$

where the last inequality uses the fact that $\|\boldsymbol{\mu}_2 - \boldsymbol{\mu}_1\|_2, \|\boldsymbol{\beta}\|_2 \lesssim \Delta$, since $\Delta = \sqrt{(\boldsymbol{\mu}_2 - \boldsymbol{\mu}_1)^\top \Omega (\boldsymbol{\mu}_2 - \boldsymbol{\mu}_1)} \geq \frac{1}{\sqrt{M}}\|\boldsymbol{\mu}_2 - \boldsymbol{\mu}_1\|_2$, and $\Delta = \sqrt{\boldsymbol{\beta}^\top \Sigma \boldsymbol{\beta}} \geq \frac{1}{\sqrt{M}}\|\boldsymbol{\beta}\|_2$.

It follows that with probability at least $1 - 6p^{-1}$,

$$|(\tilde{\boldsymbol{\beta}} - \boldsymbol{\beta})^\top \Sigma (\tilde{\boldsymbol{\beta}} - \boldsymbol{\beta})| \lesssim \sqrt{\frac{s \log p}{n}} \Delta \|\tilde{\boldsymbol{\beta}} - \boldsymbol{\beta}\|_2^2 + \frac{s \log p}{n} \|\tilde{\boldsymbol{\beta}} - \boldsymbol{\beta}\|_2^2 + \sqrt{\frac{s \log p}{n}} \Delta^2 \|\tilde{\boldsymbol{\beta}} - \boldsymbol{\beta}\|_2$$

$$+ \frac{s \log p}{n} \Delta \|\tilde{\boldsymbol{\beta}} - \boldsymbol{\beta}\|_2 + \sqrt{\frac{s \log p}{n}} \cdot \|\tilde{\boldsymbol{\beta}} - \boldsymbol{\beta}\|_2^2$$

$$+ \sqrt{\frac{s \log p}{n}} \Delta \|\tilde{\boldsymbol{\beta}} - \boldsymbol{\beta}\|_2 + \|\tilde{\boldsymbol{\beta}} - \boldsymbol{\beta}\|_2 \cdot \sqrt{\frac{s \log p}{n}}$$

$$\lesssim \sqrt{\frac{s \log p}{n}} \Delta \|\tilde{\boldsymbol{\beta}} - \boldsymbol{\beta}\|_2^2 + \sqrt{\frac{s \log p}{n}} \Delta^2 \|\tilde{\boldsymbol{\beta}} - \boldsymbol{\beta}\|_2,$$

where the last inequality uses the fact that $\Delta \geq M_{n,p} \geq c_L > 0$.

On the other hand, since

$$|(\tilde{\boldsymbol{\beta}} - \boldsymbol{\beta})^\top \Sigma (\tilde{\boldsymbol{\beta}} - \boldsymbol{\beta})| \geq \lambda_{\min}(\Sigma) \|\tilde{\boldsymbol{\beta}} - \boldsymbol{\beta}\|_2^2 \geq \frac{1}{M} \|\tilde{\boldsymbol{\beta}} - \boldsymbol{\beta}\|_2^2.$$

We then have, with probability at least $1 - 6p^{-1}$,

$$\|\tilde{\boldsymbol{\beta}} - \boldsymbol{\beta}\|_2 \lesssim \sqrt{\frac{s \log p}{n}} \left( \Delta \|\tilde{\boldsymbol{\beta}} - \boldsymbol{\beta}\|_2 + \Delta^2 \right),$$

Assuming $M_{n,p} \sqrt{\frac{s \log p}{n}} = o(1)$, which implies $\Delta \sqrt{\frac{s \log p}{n}} = o(1)$, then we have

$$\|\tilde{\boldsymbol{\beta}} - \boldsymbol{\beta}\|_2 \lesssim \frac{\Delta^2 \sqrt{\frac{s \log p}{n}}}{1 - \Delta \sqrt{\frac{s \log p}{n}}}.$$

Since $\|\tilde{\boldsymbol{\beta}}\|_1 \leq \|\boldsymbol{\beta}\|_1$ and combining with Lemma 5, we then have with probability at least $1 - 7p^{-1}$,

$$\left| \frac{\hat{\Delta}^2 - \Delta^2}{\Delta^2} \right| \leq \frac{|\tilde{\boldsymbol{\beta}}^\top (\boldsymbol{\delta} - \hat{\boldsymbol{\delta}})| + |\boldsymbol{\delta}^\top (\tilde{\boldsymbol{\beta}} - \boldsymbol{\beta})|}{\Delta^2} \leq \frac{\|\boldsymbol{\beta}\|_1 \cdot \|\boldsymbol{\delta} - \hat{\boldsymbol{\delta}}\|_\infty + \|\boldsymbol{\delta}\|_2 \cdot \|\boldsymbol{\beta} - \tilde{\boldsymbol{\beta}}\|_2}{\Delta^2}$$

$$\leq \frac{\sqrt{s} \cdot \|\boldsymbol{\beta}\|_2 \cdot \|\boldsymbol{\delta} - \hat{\boldsymbol{\delta}}\|_\infty + \|\boldsymbol{\delta}\|_2 \cdot \|\boldsymbol{\beta} - \tilde{\boldsymbol{\beta}}\|_2}{\Delta^2}$$

$$\lesssim \frac{\sqrt{s} \cdot \Delta \sqrt{\frac{\log p}{n}} + \Delta \cdot \frac{\Delta^2 \sqrt{\frac{s \log p}{n}}}{1 - \Delta \sqrt{\frac{s \log p}{n}}}}{\Delta^2} = o(1),$$

given $\Delta \geq c_L$ and $\Delta \sqrt{\frac{s \log p}{n}} = o(1)$.

Secondly, let's proceed to showing the accuracy of $\hat{\boldsymbol{\beta}}_{\text{AdaLDA}}$. We use $\hat{\boldsymbol{\beta}}$ to denote $\hat{\boldsymbol{\beta}}_{\text{AdaLDA}}$ in this subsection for simplicity. By Lemma 1, $\boldsymbol{\beta}$ lies in the feasible set of (9), so $\|\hat{\boldsymbol{\beta}}\|_1 \leq$ $\|\boldsymbol{\beta}\|_1$. By a similar argument as in (19), we have that with probability at least $1 - 3p^{-1}$,

$$
\begin{aligned}
&|(\hat{\boldsymbol{\beta}} - \boldsymbol{\beta})^\top \Sigma (\hat{\boldsymbol{\beta}} - \boldsymbol{\beta})| \\
&\leq |(\hat{\boldsymbol{\beta}} - \boldsymbol{\beta})^\top (\hat{\Sigma}\hat{\boldsymbol{\beta}} - \hat{\boldsymbol{\delta}})| + |(\hat{\boldsymbol{\beta}} - \boldsymbol{\beta})^\top (\hat{\Sigma} - \Sigma)\hat{\boldsymbol{\beta}})| + |(\hat{\boldsymbol{\beta}} - \boldsymbol{\beta})^\top (\boldsymbol{\delta} - \hat{\boldsymbol{\delta}})| \\
&\lesssim \sqrt{s} \|\hat{\boldsymbol{\beta}} - \boldsymbol{\beta}\|_2 \cdot \|\hat{\Sigma}\hat{\boldsymbol{\beta}} - \hat{\boldsymbol{\delta}}\|_\infty + \|\hat{\boldsymbol{\beta}} - \boldsymbol{\beta}\|_2 \cdot \sqrt{\frac{s \log p}{n}} \cdot \|\boldsymbol{\beta} - \hat{\boldsymbol{\beta}}\|_2 \\
&\quad + \|\boldsymbol{\beta} - \hat{\boldsymbol{\beta}}\|_2 \sqrt{\frac{s \log p}{n}} \cdot \|\boldsymbol{\beta}\|_2 + \|\boldsymbol{\beta} - \hat{\boldsymbol{\beta}}\|_2 \sqrt{\frac{s \log p}{n}}.
\end{aligned} \quad (20)
$$

Now since we have $|\frac{\hat{\Delta}^2 - \Delta^2}{\Delta^2}| = o(1)$ with probability at least $1 - 7p^{-1}$, this implies with probability at least $1 - 10p^{-1}$,

$$\|\hat{\Sigma}\hat{\boldsymbol{\beta}} - \hat{\boldsymbol{\delta}}\|_\infty \leq \sqrt{\frac{\log p}{n}} \cdot \sqrt{\hat{\sigma}_{jj}\hat{\Delta}^2 + 2\hat{\sigma}_{jj}} \lesssim \Delta \sqrt{\frac{\log p}{n}}.$$

Then using the fact $|(\tilde{\boldsymbol{\beta}} - \boldsymbol{\beta})^\top \Sigma (\tilde{\boldsymbol{\beta}} - \boldsymbol{\beta})| \geq \lambda_{\min}(\Sigma) \|\tilde{\boldsymbol{\beta}} - \boldsymbol{\beta}\|_2^2$ again, we have with probability at least $1 - 10p^{-1}$,

$$\|\hat{\boldsymbol{\beta}} - \boldsymbol{\beta}\|_2^2 \lesssim \Delta \sqrt{\frac{s \log p}{n}} \cdot \|\hat{\boldsymbol{\beta}} - \boldsymbol{\beta}\|_2 + \sqrt{\frac{s \log p}{n}} \cdot \|\hat{\boldsymbol{\beta}} - \boldsymbol{\beta}\|_2^2.$$

This implies that there exists some constant $C > 0$, such that with probability at least $1 - 10p^{-1}$,

$$\|\hat{\boldsymbol{\beta}}_{\text{AdaLDA}} - \boldsymbol{\beta}\|_2 \leq C\Delta \cdot \sqrt{\frac{s \log p}{n}}.$$

In addition, since $\|\hat{\boldsymbol{\beta}}_{\text{AdaLDA}}\|_1 \leq \|\boldsymbol{\beta}\|_1 \leq \sqrt{p}\|\boldsymbol{\beta}\|_2 \leq \sqrt{pM} \cdot \Delta$, we then have

$$
\begin{aligned}
&\mathbb{E}[\|\hat{\boldsymbol{\beta}}_{\text{AdaLDA}} - \boldsymbol{\beta}\|_2] \\
&\leq \mathbb{E}[\|\hat{\boldsymbol{\beta}}_{\text{AdaLDA}} - \boldsymbol{\beta}\|_2 \cdot 1_{\{\|\hat{\boldsymbol{\beta}}_{\text{AdaLDA}} - \boldsymbol{\beta}\|_2 > C\Delta \cdot \sqrt{\frac{s \log p}{n}}\}}] + \mathbb{E}[\|\hat{\boldsymbol{\beta}}_{\text{AdaLDA}} - \boldsymbol{\beta}\|_2 \cdot 1_{\{\|\hat{\boldsymbol{\beta}}_{\text{AdaLDA}} - \boldsymbol{\beta}\|_2 \leq C\Delta \cdot \sqrt{\frac{s \log p}{n}}\}}] \\
&\leq \sqrt{pM} \cdot \Delta \cdot 10p^{-1} + C\Delta \cdot \sqrt{\frac{s \log p}{n}} \lesssim \Delta \cdot \sqrt{\frac{s \log p}{n}} \lesssim M_{n,p} \cdot \sqrt{\frac{s \log p}{n}}.
\end{aligned}
$$

## 5.2 Proofs of Theorem 3

For a vector $\boldsymbol{x} \in \mathbb{R}^p$, we define $\|\boldsymbol{x}\|_{2,s} = \sup_{\|\boldsymbol{y}\|_2=1, \boldsymbol{y} \in \Gamma(s)} |\boldsymbol{x}^\top \boldsymbol{y}|$. We start with the following lemma.

**Lemma 7.** *For two vectors $\boldsymbol{\gamma}$ and $\hat{\boldsymbol{\gamma}}$, if $\|\boldsymbol{\gamma} - \hat{\boldsymbol{\gamma}}\|_2 = o(1)$ as $n \to \infty$, and $\|\boldsymbol{\gamma}\|_2 \geq c$ for some constant $c > 0$, then when $n \to \infty$,*

$$\|\boldsymbol{\gamma}\|_2 \cdot \|\hat{\boldsymbol{\gamma}}\|_2 - \boldsymbol{\gamma}^\top \hat{\boldsymbol{\gamma}} \asymp \|\boldsymbol{\gamma} - \hat{\boldsymbol{\gamma}}\|_2^2.$$

We postpone the proof of Lemma 7 to Section A.6 in the supplement, and continue the proof of Theorem 3.

Let $\delta_n = \|\hat{\boldsymbol{\beta}} - \boldsymbol{\beta}\|_2 \vee \|\hat{\boldsymbol{\mu}}_1 - \boldsymbol{\mu}_1\|_{2,s} \vee \|\hat{\boldsymbol{\mu}}_2 - \boldsymbol{\mu}_2\|_{2,s}$. We are going to show

$$R_{\boldsymbol{\theta}}(\hat{C}) - R_{\mathrm{opt}}(\boldsymbol{\theta}) \lesssim e^{-\Delta^2/8} \cdot \Delta \cdot \delta_n^2.$$

Given the estimators $\hat{\omega}, \hat{\boldsymbol{\mu}}_k$, and $\hat{\boldsymbol{\beta}}$, the sample $\boldsymbol{Z}$ is classified as

$$\hat{C}(\boldsymbol{Z}) = \begin{cases} 1, & (\boldsymbol{Z} - (\hat{\boldsymbol{\mu}}_1 + \hat{\boldsymbol{\mu}}_2)/2)^\top \hat{\boldsymbol{\beta}} \geq 0 \\ 2, & (\boldsymbol{Z} - (\hat{\boldsymbol{\mu}}_1 + \hat{\boldsymbol{\mu}}_2)/2)^\top \hat{\boldsymbol{\beta}} < 0. \end{cases}$$

Let $\hat{\Delta} = \sqrt{\hat{\boldsymbol{\beta}}^\top \Sigma \hat{\boldsymbol{\beta}}}$ and $\hat{\boldsymbol{\mu}} = \frac{\hat{\boldsymbol{\mu}}_1 + \hat{\boldsymbol{\mu}}_2}{2}$. The misclassification error is

$$R_{\boldsymbol{\theta}}(\hat{C}) = \frac{1}{2}\Phi\Big(-\frac{(\hat{\boldsymbol{\mu}} - \boldsymbol{\mu}_1)^\top \hat{\boldsymbol{\beta}}}{\hat{\Delta}}\Big) + \frac{1}{2}\bar{\Phi}\Big(-\frac{(\hat{\boldsymbol{\mu}} - \boldsymbol{\mu}_2)^\top \hat{\boldsymbol{\beta}}}{\hat{\Delta}}\Big),$$

with $R_{\mathrm{opt}}(\boldsymbol{\theta}) = \frac{1}{2}\Phi\big(-\Delta/2\big) + \frac{1}{2}\bar{\Phi}\big(\Delta/2\big)$. Define an intermediate quantity

$$R^* = \frac{1}{2}\Phi\Big(-\frac{\boldsymbol{\delta}^\top \hat{\boldsymbol{\beta}}/2}{\hat{\Delta}}\Big) + \frac{1}{2}\bar{\Phi}\Big(\frac{\boldsymbol{\delta}^\top \hat{\boldsymbol{\beta}}/2}{\hat{\Delta}}\Big).$$

We first show that $R^* - R_{\mathrm{opt}}(\boldsymbol{\theta}) \lesssim e^{-\Delta^2/8} \cdot \Delta \cdot \delta_n^2$. Applying Taylor's expansion to the two terms in $R^*$ at $\frac{\Delta}{2}$ and $-\frac{\Delta}{2}$ respectively, we obtain

$$R^* - R_{\mathrm{opt}}(\boldsymbol{\theta}) = \frac{1}{2}\Big(\frac{\Delta}{2} - \frac{\boldsymbol{\delta}^\top \hat{\boldsymbol{\beta}}}{2\hat{\Delta}}\Big)\Phi'\Big(\frac{\Delta}{2}\Big) + \frac{1}{2}\Big(-\frac{\boldsymbol{\delta}^\top \hat{\boldsymbol{\beta}}}{2\hat{\Delta}} + \frac{\Delta}{2}\Big)\Phi'\Big(-\frac{\Delta}{2}\Big) + O\Big(e^{-\Delta^2/8}\frac{1}{\Delta} \cdot \delta_n^4\Big), \tag{21}$$

In fact, the remaining term can be written as

$$\frac{1}{2}\Big(\frac{\boldsymbol{\delta}^\top \hat{\boldsymbol{\beta}}}{2\hat{\Delta}} - \frac{\Delta}{2}\Big)^2 \Phi''(t_{1,n}) + \Big(\frac{\boldsymbol{\delta}^\top \hat{\boldsymbol{\beta}}}{2\hat{\Delta}} - \frac{\Delta}{2}\Big)^2 \Phi''(t_{2,n}),$$

where $t_{1,n}, t_{2,n}$ are some constants satisfying $|t_{1,n}|, |t_{2,n}|$ are between $\frac{\Delta}{2}$ and $\frac{\boldsymbol{\delta}^\top \hat{\boldsymbol{\beta}}/2}{\hat{\Delta}}$.

Therefore, the remaining term can be bounded by using the facts that

$$\Big|\frac{\boldsymbol{\delta}^\top \hat{\boldsymbol{\beta}}}{2\hat{\Delta}} - \frac{\Delta}{2}\Big| = O\Big(\frac{1}{\Delta}\delta_n^2\Big), \text{ and } \Phi''(t_n) = O(e^{-\Delta^2/8}\Delta),$$

for $|t_n|$ is between $\frac{\Delta}{2}$ and $\frac{\boldsymbol{\delta}^\top \hat{\boldsymbol{\beta}}/2}{\hat{\Delta}}$.

In fact, for the first term, we can obtain this inequality by letting $\boldsymbol{\gamma} = \Sigma^{1/2}\boldsymbol{\beta}$ and $\hat{\boldsymbol{\gamma}} = \Sigma^{1/2}\hat{\boldsymbol{\beta}}$ in Lemma 7. Then

$$\Big|\Delta - \frac{\boldsymbol{\delta}^\top \hat{\boldsymbol{\beta}}}{\hat{\Delta}}\Big| = \Big|\|\boldsymbol{\gamma}\|_2 - \frac{\boldsymbol{\gamma}^\top \hat{\boldsymbol{\gamma}}}{\|\hat{\boldsymbol{\gamma}}\|_2}\Big| = \Big|\frac{\|\boldsymbol{\gamma}\|_2\|\hat{\boldsymbol{\gamma}}\|_2 - \boldsymbol{\gamma}^\top \hat{\boldsymbol{\gamma}}}{\|\hat{\boldsymbol{\gamma}}\|_2}\Big| \lesssim \frac{1}{\Delta}\|\hat{\boldsymbol{\gamma}} - \boldsymbol{\gamma}\|_2^2 \lesssim \frac{1}{\Delta}\delta_n^2.$$

In addition, since as $\delta_n \to 0$, $\frac{(\boldsymbol{\delta}^*)^\top \hat{\boldsymbol{\beta}}/2}{\hat{\Delta}} \to \frac{\Delta}{2}$, we then have $|\Phi''(t_n)| \asymp \Delta \cdot e^{-\frac{(\Delta/2)^2}{2}} = \Delta \cdot e^{-\Delta^2/8}$.

Then (21) can be further expanded such that

$$R^* - R_{\mathrm{opt}}(\boldsymbol{\theta}) \asymp \Big(-\frac{\boldsymbol{\delta}^\top \hat{\boldsymbol{\beta}}}{2\hat{\Delta}} + \frac{\Delta}{2}\Big) e^{-\frac{1}{2}(\frac{\Delta}{2})^2} + \Big(-\frac{\boldsymbol{\delta}^\top \hat{\boldsymbol{\beta}}}{2\hat{\Delta}} + \frac{\Delta}{2}\Big) e^{-\frac{1}{2}(-\frac{\Delta}{2})^2} + O\Big(e^{-\Delta^2/8} \frac{1}{\Delta} \cdot \delta_n^4\Big)$$

$$= \exp\Big(-\frac{\Delta^2}{8}\Big) \cdot \Big(-\frac{\boldsymbol{\delta}^\top \hat{\boldsymbol{\beta}}}{\hat{\Delta}} + \Delta\Big) + O\Big(e^{-\Delta^2/8} \frac{1}{\Delta} \cdot \delta_n^4\Big)$$

$$\lesssim e^{-\Delta^2/8} \cdot |\frac{\boldsymbol{\delta}^\top \hat{\boldsymbol{\beta}}}{\hat{\Delta}} - \Delta| + O\Big(e^{-\Delta^2/8} \frac{1}{\Delta} \cdot \delta_n^4\Big) \lesssim e^{-\Delta^2/8} \cdot \delta_n^2.$$

Eventually we obtain $R^* - R_{\mathrm{opt}}(\boldsymbol{\theta}) \lesssim e^{-\Delta^2/8} \Delta \cdot \delta_n^2$.

To upper bound $R_{\boldsymbol{\theta}}(\hat{C}) - R^*$, applying Taylor's expansion to $R_{\boldsymbol{\theta}}(\hat{C})$,

$$R_{\boldsymbol{\theta}}(\hat{C}) = \frac{1}{2}\bigg\{\Phi\Big(\frac{\boldsymbol{\delta}^\top \hat{\boldsymbol{\beta}}/2}{\hat{\Delta}}\Big) + \frac{(\hat{\boldsymbol{\mu}} - \boldsymbol{\mu}_1)^\top \hat{\boldsymbol{\beta}} - \boldsymbol{\delta}^\top \hat{\boldsymbol{\beta}}/2}{\hat{\Delta}} \Phi'\Big(\frac{\boldsymbol{\delta}^\top \hat{\boldsymbol{\beta}}/2}{\hat{\Delta}}\Big) + O\Big(e^{-\Delta^2/8} \Delta \cdot \delta_n^2\Big)\bigg\}$$

$$- \frac{1}{2}\bigg\{\bar{\Phi}\Big(\frac{-\boldsymbol{\delta}^\top \hat{\boldsymbol{\beta}}/2}{\hat{\Delta}}\Big) + \frac{(\hat{\boldsymbol{\mu}} - \boldsymbol{\mu}_2)^\top \hat{\boldsymbol{\beta}} + \boldsymbol{\delta}^\top \hat{\boldsymbol{\beta}}/2}{\hat{\Delta}} \Phi'\Big(-\frac{\boldsymbol{\delta}^\top \hat{\boldsymbol{\beta}}/2}{\hat{\Delta}}\Big) + O\Big(e^{-\Delta^2/8} \Delta \cdot \delta_n^2\Big)\bigg\},$$

where the remaining term can be obtained similarly as (21) by using the fact

$$\Big|\frac{(\hat{\boldsymbol{\mu}} - \boldsymbol{\mu}_1)^\top \hat{\boldsymbol{\beta}} - \boldsymbol{\delta}^\top \hat{\boldsymbol{\beta}}/2}{\hat{\Delta}}\Big| = O(\delta_n) \text{ and } |\Phi''(\cdot)| = O(e^{-\Delta^2/8}\Delta).$$

In fact, when $|\hat{\Delta} - \Delta| \leq |\sqrt{(\hat{\boldsymbol{\beta}} - \boldsymbol{\beta})\Sigma(\hat{\boldsymbol{\beta}} - \boldsymbol{\beta})}| \lesssim \|\hat{\boldsymbol{\beta}} - \boldsymbol{\beta}\|_2 \lesssim \delta_n = o(1)$, we have

$$\Big|\frac{(\hat{\boldsymbol{\mu}} - \boldsymbol{\mu}_1)^\top \hat{\boldsymbol{\beta}} - \boldsymbol{\delta}^\top \hat{\boldsymbol{\beta}}/2}{\hat{\Delta}}\Big| \leq \frac{1}{2\Delta}|(\hat{\boldsymbol{\mu}}_2 - \hat{\boldsymbol{\mu}}_1 - \boldsymbol{\mu}_2 + \boldsymbol{\mu}_2)\hat{\boldsymbol{\beta}}| \lesssim \delta_n.$$

This leads to

$$|R_{\boldsymbol{\theta}}(\hat{C}) - R^*| \lesssim \Big|\frac{\boldsymbol{\delta}^\top \hat{\boldsymbol{\beta}}/2 - (\hat{\boldsymbol{\mu}} - \boldsymbol{\mu}_1)^\top \hat{\boldsymbol{\beta}}}{\hat{\Delta}} \Phi'\Big(\frac{\boldsymbol{\delta}^\top \hat{\boldsymbol{\beta}}/2}{\hat{\Delta}}\Big)$$

$$+ \frac{\boldsymbol{\delta}^\top \hat{\boldsymbol{\beta}}/2 + (\hat{\boldsymbol{\mu}} - \boldsymbol{\mu}_2)^\top \hat{\boldsymbol{\beta}}}{\hat{\Delta}} \Phi'\Big(-\frac{\boldsymbol{\delta}^\top \hat{\boldsymbol{\beta}}/2}{\hat{\Delta}}\Big) + O\Big(e^{-\Delta^2/8} \Delta \cdot \delta_n^2\Big)\Big|$$

$$= \Big|\frac{\boldsymbol{\delta}^\top \hat{\boldsymbol{\beta}}/2 - (\hat{\boldsymbol{\mu}} - \boldsymbol{\mu}_1)^\top \hat{\boldsymbol{\beta}}}{\hat{\Delta}} e^{-\frac{1}{2}\{\frac{\boldsymbol{\delta}^\top \hat{\boldsymbol{\beta}}/2}{\hat{\Delta}}\}^2}$$

$$+ \frac{\boldsymbol{\delta}^\top \hat{\boldsymbol{\beta}}/2 + (\hat{\boldsymbol{\mu}} - \boldsymbol{\mu}_2)^\top \hat{\boldsymbol{\beta}}}{\hat{\Delta}} e^{-\frac{1}{2}\{\frac{\boldsymbol{\delta}^\top \hat{\boldsymbol{\beta}}/2}{\hat{\Delta}}\}^2} + O\Big(e^{-\Delta^2/8} \Delta \cdot \delta_n^2\Big)\Big|.$$

Since

$$\boldsymbol{\delta}/2 - (\hat{\boldsymbol{\mu}} - \boldsymbol{\mu}_1) + \boldsymbol{\delta}/2 + (\hat{\boldsymbol{\mu}} - \boldsymbol{\mu}_2) = \boldsymbol{\delta} - (\boldsymbol{\mu}_2 - \boldsymbol{\mu}_1) = 0,$$

then it follows that

$$|R_{\boldsymbol{\theta}}(\hat{C}) - R^*| \lesssim e^{-\Delta^2/8} \Delta \cdot \delta_n^2.$$

Combining the pieces, we obtain

$$R_{\boldsymbol{\theta}}(\hat{C}) - R_{\text{opt}}(\boldsymbol{\theta}) \lesssim e^{-\Delta^2/8} \cdot \Delta \cdot \delta_n^2.$$

Finally, by Lemma 5 and the derivation in Theorem 2, with probability at least $1-12p^{-1}$, $\delta_n \lesssim M_{n,p}\sqrt{\frac{s\log p}{n}}$. In addition, $\Delta \in [M_{n,p}, 3M_{n,p}]$, we then have with probability at least $1 - 12p^{-1}$,

$$R_{\boldsymbol{\theta}}(\hat{C}) - R_{\text{opt}}(\boldsymbol{\theta}) \lesssim e^{-M_{n,p}^2/8} \cdot M_{n,p}^3 \cdot \frac{s\log p}{n}.$$

Now we consider the two cases. On the one hand, when $M_{n,p}$ is bounded by $C_b$, we have

$$R_{\boldsymbol{\theta}}(\hat{C}) - R_{\text{opt}}(\boldsymbol{\theta}) \lesssim e^{-M_{n,p}^2/8} \cdot \frac{s\log p}{n}.$$

On the other hand, when $M_{n,p} \to \infty$ as $n$ grows,

$$R_{\boldsymbol{\theta}}(\hat{C}) - R_{\text{opt}}(\boldsymbol{\theta}) \lesssim e^{-(\frac{1}{8} - \frac{3\log M_{n,p}}{M_{n,p}^2})M_{n,p}^2} \cdot M_{n,p}^3 \cdot \frac{s\log p}{n},$$

where $\frac{3\log M_{n,p}}{M_{n,p}^2}$ is an $o(1)$ term as $n \to \infty$.

### 5.3 Proofs of Theorems 4 and 5

We proceed to proving Theorems 4 and 5 under the event $\{c_1 n_0 \leq n^*_{\min}(S) \leq c_2 n_0\}$ that happens with probability at least $1 - p^{-1}$. The results then rely on the following lemma.

**Lemma 8.** *Consider the MCR model and assume that $\hat{\boldsymbol{\mu}}$, $\hat{\Sigma}$ are the generalized sample mean and sample covariance matrix respectively. If $c_1 n_0 \leq n^*_{\min}(S) \leq c_2 n_0$. then with probability at least $1 - p^{-1}$,*

$$\sup_{\boldsymbol{u}\in\Gamma(s)} \boldsymbol{u}^\top(\hat{\boldsymbol{\mu}} - \boldsymbol{\mu}) \lesssim \sqrt{\frac{s\log p}{n_0}};$$

$$\sup_{\boldsymbol{u},\boldsymbol{v}\in\Gamma(s)} \boldsymbol{u}^\top(\hat{\Sigma} - \Sigma)\boldsymbol{v} \lesssim \sqrt{\frac{s\log p}{n_0}}.$$

Given Lemma 8, the derivation of Theorems 4 is very similar to the case with AdaLDA in Section 5.1, and 5 can be derived from Theorem 4 by using the same logic as in Section 5.2, and thus are omitted.

## 5.4 Proofs of the minimax lower bound results (Theorems 6 and 7)

In this section we are going to prove Theorems 6 and 7. We start with providing lemmas that will be used in the proof.

### 5.4.1 Auxiliary lemmas

The proof of Theorem 6 relies on the following Fano's Lemma.

**Lemma 9** (Tsybakov (2009)). *Suppose $\Theta_p$ is a parameter space consisting of $M$ parameters $\boldsymbol{\theta}_0, \boldsymbol{\theta}_1, ..., \boldsymbol{\theta}_M \in \Theta_p$ for some $M > 0$, and $d(\cdot, \cdot) : \Theta_p \times \Theta_p \to \mathbb{R}^+$ is some distance. Denote $\mathbb{P}_{\boldsymbol{\theta}}$ to be some probability measure parametrized by $\boldsymbol{\theta}$. If for some constants $\alpha \in (0, 1/8), \gamma > 0$, $KL(\mathbb{P}_{\boldsymbol{\theta}_i}, \mathbb{P}_{\boldsymbol{\theta}_0}) \leq \alpha \log M/n$ for all $1 \leq i \leq M$, and $d(\boldsymbol{\theta}_i, \boldsymbol{\theta}_j) \geq \gamma$ for all $0 \leq i \neq j \leq M$, then*

$$\inf_{\hat{\boldsymbol{\theta}}} \sup_{i \in [M]} \mathbb{E}_{\boldsymbol{\theta}_i}[d_{\boldsymbol{\theta}_i}(\hat{\boldsymbol{\theta}}, \boldsymbol{\theta}_i)] \gtrsim \gamma.$$

The proof of Theorem 7, however, is not straightforward, since the excess risk $R_{\boldsymbol{\theta}}(\hat{C}) - R_{opt}(\boldsymbol{\theta})$ is not a distance as required in Lemma 9. The key step in our proof of Theorem 7 is to reduce the excess risk $R_{\boldsymbol{\theta}}(\hat{C}) - R_{opt}(\boldsymbol{\theta})$ to $L_{\boldsymbol{\theta}}(\hat{C})$, defined in (16).

The following lemma suggests that it suffices to provide a lower bound for $L_{\boldsymbol{\theta}}(\hat{C})$, and $L_{\boldsymbol{\theta}}(\hat{C})$ satisfies an approximate triangle inequality (Lemma 4).

Although $L_{\boldsymbol{\theta}}(\hat{C})$ is not a distance function and does not satisfy an exact triangle inequality, the following lemma provides a variant of Fano's Lemma.

**Lemma 10** (Tsybakov (2009)). *Let $M \geq 0$ and $\boldsymbol{\theta}_0, \boldsymbol{\theta}_1, ..., \boldsymbol{\theta}_M \in \Theta_p$. For some constants $\alpha_0 \in (0, 1/8], \gamma > 0$, and any classifier $\hat{C}$, if $KL(\mathbb{P}_{\boldsymbol{\theta}_i}, \mathbb{P}_{\boldsymbol{\theta}_0}) \leq \alpha_0 \log M/n$ for all $1 \leq i \leq M$, and $L_{\boldsymbol{\theta}_i}(\hat{C}) < \gamma$ implies $L_{\boldsymbol{\theta}_j}(\hat{C}) \geq \gamma$ for all $0 \leq i \neq j \leq M$, then*

$$\inf_{\hat{C}} \sup_{i \in [M]} \mathbb{P}_{\boldsymbol{\theta}_i}(L_{\boldsymbol{\theta}_i}(\hat{C}) \geq \gamma) \geq \frac{\sqrt{M}}{\sqrt{M}+1}\left(1 - 2\alpha_0 - \sqrt{\frac{2\alpha_0}{\log M}}\right).$$

**Lemma 11** (Tsybakov (2009)). *Define $\mathcal{A}_{p,s} = \{\boldsymbol{u} : \boldsymbol{u} \in \{0,1\}^p, \|\boldsymbol{u}\|_0 \leq s\}$. If $p \geq 4s$, then there exists a subset $\{\boldsymbol{u}_0, \boldsymbol{u}_1, ..., \boldsymbol{u}_M\} \subset \mathcal{A}_{p,s}$ such that $\boldsymbol{u}_0 = \{0, ..., 0\}^\top$, $\rho_H(\boldsymbol{u}_i, \boldsymbol{u}_j) \geq s/2$ and $\log(M + 1) \geq \frac{s}{5} \log(\frac{p}{s})$, where $\rho_H$ denotes the Hamming distance.*

### 5.4.2 Proof of Theorem 6

In this section we prove the lower bound of estimation of $\boldsymbol{\beta}$. First we construct a subset of the parameter space $\Theta$ that characterizes the hardness of the problem. By Lemma 11, there exist $\boldsymbol{u}_0, \boldsymbol{u}_1, ..., \boldsymbol{u}_M \in \mathcal{A}_{p,s} = \{\boldsymbol{u} \in \{0,1\}^p : \|\boldsymbol{u}\|_0 \leq s\}$, such that $\rho_H(\boldsymbol{u}_i, \boldsymbol{u}_j) > s/2$ and $\log(M+1) \geq \frac{s}{5} \log(\frac{p}{s})$, denote this collection of $\boldsymbol{u}_i$ by $\tilde{\mathcal{A}}_{p,s}$. In addition, denote $\boldsymbol{u}_0 = \boldsymbol{0}_p$.

Since $\frac{\log p}{\log(p/s)} = O(1)$, so for sufficiently large $p$, we have $s < p/2$. Define $\boldsymbol{b}_0$ be the $p$-dimensional vector with the last $s$ entries being $\frac{M_{n,p}}{\sqrt{s}}$ and the rest being 0, so we have $\|\boldsymbol{b}_0\|_2 = M_{n,p}$. Let $r = \lceil p/2 \rceil$. For $\boldsymbol{u} \in \tilde{\mathcal{A}}_{p,s} = \{\boldsymbol{u}_0, \boldsymbol{u}_1, ..., \boldsymbol{u}_M\}$, let $B_{\boldsymbol{u}}$ be the $p \times p$ symmetric matrix whose $i$-th row and column are both $\epsilon \cdot u_i \cdot \frac{\boldsymbol{b}_0}{M_{n,p}}$ for $i \in \{1, ..., r\}$, where $\epsilon$ is to be determined later. The parameter set we considered is

$$\Theta_0 = \{\boldsymbol{\theta} = (\boldsymbol{\mu}_1, \boldsymbol{\mu}_2, \Sigma) : \boldsymbol{\mu}_1 = \boldsymbol{b}_0, \boldsymbol{\mu}_2 = -\boldsymbol{b}_0, \Sigma = (I_p + B_{\boldsymbol{u}})^{-1}; \boldsymbol{u} \in \tilde{\mathcal{A}}_{p,s} \cup \{\boldsymbol{0}_p\}\}.$$

For a given $\boldsymbol{u}$, the corresponding discriminating direction is $\boldsymbol{\beta}_{\boldsymbol{u}} = -2(I_p + B_{\boldsymbol{u}})\boldsymbol{b}_0$, which implies

$$\|\boldsymbol{\beta}_{\boldsymbol{u}} - \boldsymbol{\beta}_{\tilde{\boldsymbol{u}}}\|_2^2 = 4\|(B_{\boldsymbol{u}} - B_{\tilde{\boldsymbol{u}}})\boldsymbol{b}_0\|_2^2 \geq 4\rho_H(\boldsymbol{u}, \tilde{\boldsymbol{u}})\epsilon^2\|\boldsymbol{b}_0\|_2^2 \geq 2sM_{n,p}^2\epsilon^2.$$

In addition, when $\|B_{\boldsymbol{u}}\|_2 = o(1)$, for sufficiently large $n$, we have $\Delta = \sqrt{4\boldsymbol{b}_0^\top (I_p + B_{\boldsymbol{u}})\boldsymbol{b}_0} \in (M_{n,p}, 3M_{n,p})$, which implies that $\Theta_0 \subset \mathcal{G}(s, M_{n,p})$.

We then proceed to bound $KL(\mathbb{P}_{\boldsymbol{\theta}_{\boldsymbol{u}_i}}, \mathbb{P}_{\boldsymbol{\theta}_{\boldsymbol{u}_0}})$ for $i \in [M]$, where $\mathbb{P}_{\boldsymbol{\theta}_{\boldsymbol{u}_i}}, \mathbb{P}_{\boldsymbol{\theta}_{\boldsymbol{u}_0}}$ denote the distributions $N_p(\boldsymbol{b}_0, (I_p + B_{\boldsymbol{u}_i})^{-1})$ and $N_p(\boldsymbol{b}_0, I_p)$ respectively. We then have

$$KL(\mathbb{P}_{\boldsymbol{\theta}_{\boldsymbol{u}_i}}, \mathbb{P}_{\boldsymbol{\theta}_{\boldsymbol{u}_0}}) = \frac{1}{2}\left[-\log|I_p + B_{\boldsymbol{u}_i}| - p + \text{tr}(I_p + B_{\boldsymbol{u}_i}))\right].$$

Note that $\frac{\boldsymbol{b}_0}{M_{n,p}}$ is a unit vector. If we take $\epsilon$ such that $\|B_{\boldsymbol{u}_i}\|_2 \leq \|B_{\boldsymbol{u}}\|_F \leq \sqrt{2s \cdot \epsilon^2} = o(1)$, and denote the eigenvalues of $I_p + B_{\boldsymbol{u}_i}$ by $1 + \Delta_{\lambda_1}, ..., 1 + \Delta_{\lambda_p}$ with $\Delta_{\lambda_j} = o(1)$. We then have

$$KL(\mathbb{P}_{\boldsymbol{\theta}_{\boldsymbol{u}_i}}, \mathbb{P}_{\boldsymbol{\theta}_{\boldsymbol{u}_0}}) = \frac{1}{2}\left[-\sum_{j=1}^{p}\log(1 + \Delta_{\lambda_j}) - p + \sum_{j=1}^{p}(1 + \Delta_{\lambda_j})\right]$$

$$\asymp \frac{1}{4}\sum_{j=1}^{p}\Delta_{\lambda_j}^2 = \frac{1}{4}\|B_{\boldsymbol{u}}\|_F^2 \leq \frac{1}{2}s\epsilon^2$$

where we use the fact that $\log(1 + x) \asymp x - \frac{x^2}{2}$ when $x = o(1)$. Now let $\epsilon = \frac{1}{5\sqrt{2}}\sqrt{\frac{\log p}{n}}$, then $KL(\mathbb{P}_{\boldsymbol{\theta}_{\boldsymbol{u}_i}}, \mathbb{P}_{\boldsymbol{\theta}_{\boldsymbol{u}_0}}) \leq \alpha \log M/n$ for $\alpha = 1/8$.

In addition, let $\gamma = \frac{1}{10}M_{n,p}\sqrt{\frac{s\log p}{n}}$, then for $0 \leq i \neq j \leq M$ and any $\hat{\boldsymbol{\beta}} \in \mathbb{R}^p$, such that $\|\hat{\boldsymbol{\beta}} - \boldsymbol{\beta}_{\boldsymbol{u}_i}\|_2 \leq \gamma$, we have

$$\|\hat{\boldsymbol{\beta}} - \boldsymbol{\beta}_{\boldsymbol{u}_j}\|_2 \geq \|\boldsymbol{\beta}_{\boldsymbol{u}_j} - \boldsymbol{\beta}_{\boldsymbol{u}_j}\|_2 - \|\hat{\boldsymbol{\beta}} - \boldsymbol{\beta}_{\boldsymbol{u}_i}\|_2 \geq \frac{1}{5}M_{n,p}\sqrt{\frac{s\log p}{n}} - \frac{1}{10}M_{n,p}\sqrt{\frac{s\log p}{n}} = \frac{1}{10}M_{n,p}\sqrt{\frac{s\log p}{n}} = \gamma.$$

Then by Fano's lemma (Lemma 9), we have $\inf_{\hat{\boldsymbol{\beta}}} \sup_{i \in [M]} \mathbb{E}\|\hat{\boldsymbol{\beta}} - \boldsymbol{\beta}_{\boldsymbol{u}_i}\|_2 \gtrsim M_{n,p}\sqrt{\frac{s\log p}{n}}$.

For the incomplete data case with $n_0 \geq 1$, we consider a special pattern of missingness $S_0$:

$$(S_0)_{ij} = 1_{\{1 \leq i \leq n_0, 1 \leq j \leq p\}} \quad \text{with probability } 1.$$

Under this missingness pattern, $n^*_{\min} = n_0$ with probability 1, and the problem essentially becomes complete data problem with $n_0$ samples, which implies

$$\inf_{\hat{\boldsymbol{\beta}}} \sup_{\substack{\boldsymbol{\theta} \in \mathcal{G}(s, M_{n,p}) \\ \boldsymbol{S} \in \boldsymbol{\Psi}(n_0; n, p)}} \mathbb{E}[\|\hat{\boldsymbol{\beta}} - \boldsymbol{\beta}\|_2] \gtrsim M_{n,p} \sqrt{\frac{s \log p}{n_0}}.$$

### 5.4.3 Proof of Theorem 7

We proceed by applying Lemma 10 to obtain the minimax lower bound for the excess misclassification error. We first construct a subset of the parameter space $\Theta$ that characterizes the hardness of the problem. Let $\boldsymbol{e}_1$ be the basis vector in the standard Euclidean space whose first entry is 1 and zero elsewhere. By Lemma 11, there exist $\boldsymbol{u}_1, ..., \boldsymbol{u}_M \in \check{\mathcal{A}}_{p,s} = \{\boldsymbol{u} \in \{0,1\}^p :, \boldsymbol{u}^\top \boldsymbol{e}_1 = 0, \|\boldsymbol{u}\|_0 = s\}$, such that $\rho_H(\boldsymbol{u}_i, \boldsymbol{u}_j) > s/2$ and $\log(M+1) \geq \frac{s}{5} \log(\frac{p-1}{s})$. Note the first entry in $\boldsymbol{u}_j$ is 0 for all $j = 1, \ldots, M$.

Define the parameter space

$$\Theta_1 = \{\boldsymbol{\theta} = (\boldsymbol{\mu}_1, \boldsymbol{\mu}_2, \Sigma) : \boldsymbol{\mu}_1 = \epsilon \boldsymbol{u} + \lambda \boldsymbol{e}_1, \boldsymbol{\mu}_2 = -\boldsymbol{\mu}_1, \Sigma = \sigma^2 I_p; \boldsymbol{u} \in \check{\mathcal{A}}_{p,s}\},$$

where $\epsilon = \sigma \sqrt{\log p / n}$, $\sigma^2 = O(1)$ and $\lambda$ is chosen to ensure $\boldsymbol{\theta} \in \mathcal{G}(s, M_{n,p})$ such that

$$(\boldsymbol{\mu}_1 - \boldsymbol{\mu}_2)^T \Sigma^{-1} (\boldsymbol{\mu}_1 - \boldsymbol{\mu}_2) = \frac{4\|\epsilon \boldsymbol{u} + \lambda \boldsymbol{e}_1\|_2^2}{\sigma^2} = M_{n,p}.$$

To apply Lemma 10, we need to verify two conditions: (i) the upper bound on the KL divergence between $\mathbb{P}_{\boldsymbol{\theta}_u}$ and $\mathbb{P}_{\boldsymbol{\theta}_v}$, and (ii) the lower bound of $L_{\boldsymbol{\theta}_u}(\hat{C}) + L_{\boldsymbol{\theta}_v}(\hat{C})$ for $\boldsymbol{u} \neq \boldsymbol{v} \in \check{\mathcal{A}}_{p,s}$.

We calculate the KL divergence first. For $\boldsymbol{u} \in \check{\mathcal{A}}_{p,s}$, denote $\boldsymbol{\mu}_u = \epsilon \boldsymbol{u} + \lambda \boldsymbol{e}_1$. For $\boldsymbol{\theta}_u = (\boldsymbol{\mu}_u, -\boldsymbol{\mu}_u, \sigma^2 I_p) \in \Theta_1$, we consider the distribution $N_p(\boldsymbol{\mu}_u, \sigma^2 I_p)$. Then, the KL divergence between $\mathbb{P}_{\boldsymbol{\theta}_u}$ and $\mathbb{P}_{\boldsymbol{\theta}_v}$ can be bounded by

$$\text{KL}(\mathbb{P}_{\boldsymbol{\theta}_u}, \mathbb{P}_{\boldsymbol{\theta}_v}) \leq \frac{1}{2} \|\boldsymbol{\mu}_u - \boldsymbol{\mu}_v\|_2^2 \leq \sigma^2 \cdot \frac{s \log p}{n}. \tag{22}$$

In addition, by applying Lemma 4, we have that for any $\boldsymbol{u}, \boldsymbol{v} \in \check{\mathcal{A}}_{p,s}$,

$$L_{\boldsymbol{\theta}_u}(\hat{C}) + L_{\boldsymbol{\theta}_v}(\hat{C}) \gtrsim \frac{1}{M_{n,p}} e^{-M_{n,p}^2/8} \sqrt{\frac{s \log p}{n}}.$$

So far we have verified the aforementioned conditions (i) and (ii). Lemma 10 immediately implies that, there is some $C_\alpha \geq 0$, such that

$$\inf_{\hat{C}} \sup_{\boldsymbol{\theta} \in \mathcal{G}(s, M_{n,p})} \mathbb{P}(L_{\boldsymbol{\theta}}(\hat{C}) \geq C_\alpha \frac{1}{M_{n,p}} e^{-M_{n,p}^2/8} \sqrt{\frac{s \log p}{n}}) \geq 1 - \alpha. \tag{23}$$

Finally combining (23) with Lemma 3, we obtain the desired lower bound for the excess misclassficiation error

$$\inf_{\hat{C}} \sup_{\boldsymbol{\theta} \in \mathcal{G}(s, M_{n,p})} \mathbb{P}(R_{\boldsymbol{\theta}}(\hat{C}) - R_{\mathrm{opt}}(\boldsymbol{\theta}) \geq C_{\alpha} \frac{1}{M_{n,p}} e^{-M_{n,p}^2/8} \frac{s \log p}{n}) \geq 1 - \alpha.$$

Under this missingness data case, we consider the same missingness pattern $S_0$ as described in Section 5.4.2 with $n_{\min} = n_0$. Then we have

$$\inf_{\hat{\boldsymbol{\beta}}} \sup_{\substack{\boldsymbol{\theta} \in \mathcal{G}(s, M_{n,p}) \\ \boldsymbol{S} \in \boldsymbol{\Psi}(n_0; n, p)}} \mathbb{P}(R_{\boldsymbol{\theta}}(\hat{C}) - R_{\mathrm{opt}}(\boldsymbol{\theta}) \geq C_{\alpha} \frac{1}{M_{n,p}} e^{-M_{n,p}^2/8} \frac{s \log p}{n_0}) \geq 1 - \alpha.$$

This implies that

1. If $M_{n,p} \leq C_b$ for some $C_b > 0$, then

$$\inf_{\hat{C}} \sup_{\substack{\boldsymbol{\theta} \in \mathcal{G}(s, M_{n,p}) \\ \mathcal{F} \in \boldsymbol{\Psi}(n_0; n, p)}} \mathbb{P}(R_{\boldsymbol{\theta}}(\hat{C}) - R_{\mathrm{opt}}(\boldsymbol{\theta}) \geq C_{\alpha} e^{-\frac{1}{8} M_{n,p}^2} \cdot \frac{s \log p}{n_0}) \geq 1 - \alpha.$$

2. If $M_{n,p} \to \infty$ as $n \to \infty$, then for any $\delta > 0$,

$$\inf_{\hat{C}} \sup_{\substack{\boldsymbol{\theta} \in \mathcal{G}(s, M_{n,p}) \\ \mathcal{F} \in \boldsymbol{\Psi}(n_0; n, p)}} \mathbb{P}(R_{\boldsymbol{\theta}}(\hat{C}) - R_{\mathrm{opt}}(\boldsymbol{\theta}) \geq C_{\alpha} e^{-(\frac{1}{8}+\delta) M_{n,p}^2} \cdot \frac{s \log p}{n_0}) \geq 1 - \alpha.$$